\documentclass[journal]{IEEEtran}
\IEEEoverridecommandlockouts
\usepackage{cite}
\usepackage{amsmath,amssymb,amsfonts}
\usepackage{algorithmic}
\usepackage{textcomp}
\usepackage{xcolor}
\usepackage{times}
\usepackage{dsfont}
\usepackage{epsfig,verbatim}
\usepackage{subfigure}
\usepackage{setspace}
\usepackage{color}
\usepackage{cite}
\usepackage{epstopdf}
\usepackage{graphics}
\usepackage{accents}
\usepackage{acronym}
\usepackage[bookmarks,colorlinks]{hyperref}
\usepackage{booktabs}
\usepackage{mathtools}
\usepackage{algorithm}
\usepackage{stfloats}
\usepackage{enumitem}
\usepackage{bbm}

\def\BibTeX{{\rm B\kern-.05em{\sc i\kern-.025em b}\kern-.08em
    T\kern-.1667em\lower.7ex\hbox{E}\kern-.125emX}}

\newtheorem{lemma}{Lemma}
\newtheorem{remark}{Remark}

\acrodef{dma}[DMA]{dynamic metasurface antenna}
\acrodef{sinr}[SINR]{signal-to-interference-and-noise ratio}
\acrodef{bs}[BS]{base station} 
\acrodef{em}[EM]{electromagnetic} 
\acrodef{mimo}[MIMO]{multiple-input multiple-output}
\acrodef{ris}[RIS]{reconfigurable intelligent surface}
\acrodef{his}[HIS]{holographic intelligent surface}
\acrodef{si}[SI]{self-interference}  
\acrodef{awgn}[AWGN]{additive white Gaussian noise} 
\acrodef{ula}[ULA]{uniform linear array}
\acrodef{upa}[UPA]{uniform planar array}
\acrodef{isac}[isac]{dual-function radar-communication}
\acrodef{his}[his]{hybrid \ac{ris}}
\acrodef{fgs}[FGS]{fast grid search}
\acrodef{agd}[AGD]{auto gradient descent}
\acrodef{rf}[RF]{radio frequency}
\acrodef{fov}[FOV]{field of view}
\acrodef{ga}[GA]{genetic algorithm}
\acrodef{sdp}[SDP]{semidefinite programming}
\acrodef{sdr}[SDR]{semidefinite relaxation}
\acrodef{isac}[ISAC]{integrated sensing and communication}
\acrodef{ad}[AD]{automatic differentiation}
\acrodef{ao}[AO]{alternating optimization}
\acrodef{his}[HIS]{holographic intelligence surface}
\acrodef{cd}[CD]{coordinate descent}
\acrodef{hisac}[HISAC]{HIS-assisted \ac{isac}}
\acrodef{abs}[ABS]{adaptive bisection searching}

\begin{document}

\title{Holographic Intelligence Surface Assisted Integrated Sensing and Communication
}

\author{  
	\IEEEauthorblockN{Zhuoyang Liu,~\IEEEmembership{Graduated Student Member,~IEEE}, Yuchen Zhang,~\IEEEmembership{Graduated Student Member,~IEEE}, \\
 Haiyang Zhang,~\IEEEmembership{Member,~IEEE}, Feng Xu,~\IEEEmembership{Senior Member,~IEEE}, 
 and Yonina C. Eldar,~\IEEEmembership{Fellow,~IEEE}\\
	} 
\thanks{
    Z. Liu and F. Xu are with the Key Lab for Information Science of Electromagnetic Wave (MoE), Fudan University, Shanghai 200433, China (e-mail: \{liuzy20; fengxu\}@fudan.edu.cn).
    
    Y. Zhang is with the National Key Laboratory of Wireless Communications, University of Electronic Science and Technology of China, Chengdu 611731, China (e-mail: yc\_zhang@std.uestc.edu.cn).
    
    H. Zhang is with the School of Communication and Information Engineering, Nanjing University of Posts and Telecommunications, Nanjing 210003, China (e-mail: haiyang.zhang@njupt.edu.cn).
    
    Y. C. Eldar is with the Faculty of Math and CS, Weizmann Institute of Science, Rehovot 7610001, Israel (e-mail: yonina.eldar@weizmann.ac.il).
    }
	
	\vspace{-1.0cm}

 }

\maketitle

\begin{abstract}
    Traditional discrete-array-based systems fail to fully exploit interactions between closely spaced antennas, resulting in inadequate utilization of the aperture resource. In this paper, we propose a holographic intelligence surface (HIS) assisted integrated sensing and communication (HISAC) system, wherein both the transmitter and receiver are fabricated using a continuous-aperture array.
    To enable tractable design and optimization, we introduce a Fourier-based continuous-discrete transformation, converting the continuous pattern design into a finite-dimensional representation in the wavenumber domain.
    Based on this transformation, we formulate a joint transmit–receive beamforming optimization problem that aims to maximize multi-target sensing performance while satisfying multi-user communication requirements.
    To solve the non-convex problem with coupled variables, an alternating optimization-based algorithm is proposed to optimize the HISAC transmit-receive beamforming in an alternate manner. Specifically, the transmit beamforming design is solved by decoupling into a series of feasibility-checking sub-problems while the receive beamforming is determined by the Rayleigh quotient-based method.
    Simulation results demonstrate the superiority of the proposed HISAC system over traditional discrete-array-based ISAC systems, achieving significantly higher sensing performance while guaranteeing predetermined communication performance.

\end{abstract}


\section{Introduction}
\label{intro}

\Acf{isac} system, which envisions simultaneous sensing and communication on a shared waveform/platform, is considered a vital solution for tackling challenges associated with spectral and/or power efficiency in 6G wireless networks \cite{2,liu2022integrated}. 
In a multi-antenna setting, joint ISAC waveform design is typically achieved through transmit beamforming design. 
This entails designing spatial beamformers judiciously to maximize sensing/communication under performance constraints in communication/sensing, thereby striking a balance between the two functionalities \cite{4,hua2023optimal,he2022energy}. Furthermore, by leveraging radar receivers, joint transmit-receive beamforming has been proposed in the literature to further enhance \ac{isac} performance \cite{wang2023ris,chen2022generalized,liu2021cramer}. 

The aforementioned beamforming designs in \ac{isac} systems typically rely on discrete antenna arrays, where the antenna spacing is constrained to half-wavelength. 
This limits the number of antennas for a given aperture, thereby resulting in insufficient exploitation of array gain for \ac{isac} performance enhancement \cite{demir2022channel,deng2023reconfigurable}.
In addition, modeling \ac{em} propagation for the discrete array usually overlooks the mutual interference between two closely spaced antennas, potentially leading to inadequate information acquisition during signal reception \cite{liu2019hybrid,gradoni2021end}. 
To overcome these limitations, the concept of \ac{his}, also as known as holographic \ac{mimo} \cite{pizzo2022fourier,pizzo2020spatially} and continuous-aperture \ac{mimo} \cite{jensen2008capacity,zhang2023pattern}, has emerged.
By forming an effectively continuous-aperture, HIS enables more complete utilization of the physical aperture of the antenna array, thereby maximizing the radiated energy in the desired direction and approaching the aperture-limited directivity. 
Unlike the discrete array consisting of multiple patch antennas with half-wavelength spacing, \ac{his} achieves a quasi-continuous-aperture array by deploying massive sub-wavelength tunable elements in a compact space \cite{zhang2022holographic}.
This architecture mitigates the element-spacing constraints inherent in discrete arrays, enabling higher array gain for the same physical aperture size \cite{deng2023reconfigurable,deng2023reconfigurable2}.
The advantages of \ac{his} include enhancing signal coverage \cite{wei2022multi}, improving communication rate \cite{hu2018beyond}, and boosting localization accuracy \cite{huang2020holographic,deng2023reconfigurable}.
However, due to its continuous nature, \ac{his} incurs significant challenges for array signal processing, making traditional beamforming techniques \cite{4,hua2023optimal,he2022energy,wang2023ris,chen2022generalized,liu2021cramer}, which are tailored for discrete arrays, inefficient or even inapplicable \cite{pizzo2022fourier,pizzo2020spatially}.
Due to these challenges, integrating \ac{his} into the \ac{isac} system, though promising for achieving a better performance trade-off between sensing and communication, is still in its infancy. 

The primary goals of this work are to: 1) design an \ac{hisac} transceiver architecture, 2) develop an \ac{hisac} transmit-receive
 beamforming framework, and 3) demonstrate the performance gain by employing \ac{his} instead of traditional discrete arrays in \ac{isac} systems.


\subsection{Prior work}

Many efforts have been dedicated to beamforming design for the \ac{isac} systems \cite{4,hua2023optimal,he2022energy,wang2023ris,pritzker2022transmit,chen2022generalized,liu2021cramer,wen2023efficient,liu2024hybrid}, including transmit beamforming and joint transmit-receive beamforming. Transmit beamforming builds on the spatial degrees of freedom (DoFs) at the transmitter, which is devised to strike performance trade-off between sensing and communication \cite{4,hua2023optimal,he2022energy}. 
A commonly adopted strategy of transmit beamforming is to modify the transmit beam by optimizing the beamformer, aiming to maximize the power sent towards the targets while fulfilling the communication performance requirement at each user \cite{4,hua2023optimal,he2022energy}.

Transmit beamforming fails to explore the spatial DoFs at the radar receiver, which holds potential to further enhance \ac{isac} performance.
Joint transmit-receive beamforming, in which the transmitter-side beamformers and radar receiver-side beamformers/filters are designed simultaneously,
has been proposed in \cite{wen2023efficient,wang2023ris,chen2022generalized,liu2021cramer,pritzker2022transmit,zhang2024robust,liu2024hybrid}. 
In \cite{pritzker2022transmit}, the authors considered transmit-receive beamforming in \ac{isac} systems with \ac{sinr} being the radar performance metric. However, without being optimized adaptively, the receive beamforming conducted at the radar receiver was solely determined by the steering vector. As a step further, \cite{zhang2024robust} proposed a design framework for \ac{isac} transmit-receive beamforming, optimizing transmit and receive beamformers alternately. 
The aforementioned works regarding \ac{isac} beamforming typically rely on a discrete array, which limits performance due to inefficient utilization of a given aperture \cite{liu2019hybrid,gradoni2021end}.

\begin{table*}[t]
\centering
\small
\caption{Summary Comparison of Related HIS / Holographic Systems}

\label{tab:compare_fourier}
\setlength{\tabcolsep}{4pt}
\renewcommand{\arraystretch}{1.15}
\begin{tabular}{lccc}
\hline
\textbf{Work} 
& \textbf{Aperture Model} 
& \textbf{System Scope} 
& \textbf{Control Variables} 
\\
\hline
Holographic MIMO \cite{cao2026hybrid}
& Discrete (element-level)
& Localization
& Phase shifts of RHS elements
\\
Circular Holographic MIMO \cite{huang2025circular}
& Discrete (element-level)
& Comm. / Throughput
& Phase shifts of RHS elements
\\
Overview of RHSs \cite{di2025reconfigurable}
& Discrete (element-level)
& ISAC
& Amplitude and phase of RHS elements
\\
Shape-adaptive RHSs \cite{jalali2025shape}
& Discrete (element-level)
& Comm. / Throughput
& Amplitude, phase, and geometry parameters
\\
RHS-assisted ISAC \cite{zhang2022holographic,zhang2023holographic}
& Discrete (element-level)
& ISAC
& Amplitude and phase of RHS elements
\\
Continuous-aperture HIS \cite{zhang2023pattern,zhu2021electromagnetic}
& Continuous
& Comm. / Throughput
& EM-consistent surface patterns
\\
\textbf{This work} 
& \textbf{Continuous} 
& \textbf{ISAC} 
& \textbf{Wavenumber-domain coefficients} 
\\
\hline
\end{tabular}
\parbox{\linewidth}{\footnotesize
\vspace{2pt}
\textit{Note:} Holographic MIMO, reconfigurable holographic surfaces (RHSs), and HISs denote related surface-based architectures. In particular, discrete-aperture HMIMO/RHS systems correspond to element-level control of amplitude and phase, whereas continuous-aperture HIS models rely on continuous electromagnetic representations or equivalent wavenumber-domain coefficients.\\
\textit{Abbreviations:} Comm. : Communication.
}
\end{table*}

With the development of continuous-aperture array technology, \ac{his} has been devised to alleviate the performance bottleneck caused by traditional discrete arrays, enabling higher array gain for a given aperture size \cite{demir2022channel,deng2023reconfigurable,liu2019hybrid,gradoni2021end,pizzo2022fourier,pizzo2020spatially}. 
The superiority of \ac{his} has been demonstrated in various systems aimed at different tasks, including communication, localization, and other integrated services \cite{deng2021reconfigurable,gong2024holographic,cao2026hybrid,huang2025circular,di2025reconfigurable,jalali2025shape,zhang2022holographic,zhang2023holographic}.
Existing holographic systems are predominantly modeled using sub-wavenumber discrete dipole representations to approximate quasi-continuous distributions \cite{di2025reconfigurable,cao2026hybrid,huang2025circular,zhang2022holographic,zhang2023holographic,jalali2025shape}. Such formulations have been widely adopted for communication-centric designs \cite{huang2025circular,cao2026hybrid,jalali2025shape} and \ac{isac} systems \cite{zhang2022holographic,zhang2023holographic}, as summarized in Table~\ref{tab:compare_fourier}. In particular, \cite{jalali2025shape} adopts a discrete element-level model and focuses on optimizing surface geometry and phase shifts for communication-oriented performance. By contrast, continuous-aperture \ac{his} models describe the surface using continuous electromagnetic fields, which provide a more fundamental characterization of the spatial DoF.

To characterize the radiation model of \ac{his}, it is necessary to design the surface pattern, known as the current density distribution, for a continuous-aperture array \cite{pizzo2020spatially,pizzo2022fourier,zhang2023pattern,jensen2008capacity}.
The authors of \cite{Sanguinetti2022wavenumber,wan2023mutual} modeled the radiation of continuous-aperture arrays and derived a closed-form expression of the capacity between two continuous volumes by utilizing the Fourier basis functions of continuous-space \ac{em} channels. 
To optimize the capacity of HIS-based communications, a pattern division multiplexing (PDM) method was developed to model the radiation for \ac{his} \cite{zhang2023pattern},  aiming to maximize the sum rate of multi-users with HIS.

\subsection{Motivation and Contribution}
{Table~\ref{tab:compare_fourier} provides a high-level comparison of representative discrete-array, holographic, and \ac{his}-based systems.}
Most existing works employing \ac{his} have focused on communication tasks \cite{zhu2021electromagnetic}, whereas the incorporation of \ac{his} into \ac{isac} systems has remained in its infancy \cite{zhang2022holographic,zhang2023holographic}.
Although the advantages of deploying sub-half-wavelength-spacing antennas for \ac{isac} systems have been initially revealed in \cite{zhang2022holographic,zhang2023holographic}, the advocated antenna array remained discrete. 
Therefore, the proposed schemes are not applicable to \ac{hisac} systems comprised of continuous-aperture arrays.

To fill the gap, this work focuses on incorporating \ac{his} into \ac{isac} systems. Specifically, we propose an \ac{isac} transceiver structure based on \ac{his}, in which both the transmitter and receiver at the \ac{bs} are implemented using continuous-aperture \acp{his}.
By generating a specific pattern of the \ac{his}, the radiation of the \ac{em} waves can be configured to support multi-target sensing and multi-user communication simultaneously. 
However, due to its continuity, the pattern of the \ac{his} has infinite dimensions, posing significant computational challenges.
To address this, we develop a continuous-discrete transformation based on the Fourier series, which maps the spatial-domain current into a finite-dimensional representation in the wavenumber domain.
Compared to \cite{zhang2023pattern,Sanguinetti2022wavenumber}, we further derive a closed-form expression for the far-field Green's function in the Fourier domain, which enables precise \ac{sinr} evaluation.
This allows the original pattern design to be reformulated as a joint transmit–receive beamforming, where the beamforming matrices in the wavenumber domain are optimized to configure the radiation beams of the \ac{his}.
We then jointly optimize the transmit-receive beamformers to maximize the minimum sensing \ac{sinr} across multiple radar targets, while ensuring that communication \ac{sinr} requirements for all users are satisfied.

The main contributions of this work are summarized as follows:
\begin{itemize}



    \item \textbf{HISAC Tranceiver}:
    To the best of our knowledge, we are the first to propose the \ac{hisac} system, where the transmit \ac{his} supports multi-user communication and illuminates multiple radar targets, and the receive \ac{his} collects the radar echoes. 
    Specifically, we study the joint design of transmit \ac{his} and receive \ac{his}, balancing the performance of sensing and communication in the proposed system. 

 
    \item \textbf{Fourier-Domain Modeling of Continuous HIS Apertures}:
    We propose a physically consistent Fourier-domain transformation that maps the continuous surface current distribution of a \ac{his} into a finite-dimensional representation in the wavenumber domain. This leads to an equivalent steering vector expression for continuous-apertures, enabling direct alignment between \ac{his}-based \ac{isac} systems and conventional discrete-array models.

    \item \textbf{Discrete-Compatible Reformulation of the HISAC Problem}:
    Leveraging the equivalent representation, we reformulate the originally infinite-dimensional beamforming problem into a finite-dimensional structure. The resulting optimization problem shares a similar mathematical form to those for discrete-array \ac{isac} systems, which allows the direct application of existing optimization techniques. To support this formulation, we derive a closed-form expression for the Fourier transform of the far-field Green’s function, facilitating precise \ac{sinr} evaluation in the \ac{hisac} problem.
 
    \item \textbf{Joint Transmit-Receive Beamforming in HISAC}:
    We develop an \ac{ao} algorithm by recasting the original problem into: transmit \ac{his} design and receive \ac{his} configuration. Due to the high complexity of the involved \ac{sdr}-based method, we propose an \ac{abs} method to accelerate the convergence in the iteration process. To address the receive \ac{his} configuration, a Rayleigh quotient-based method is proposed to obtain a closed-form solution.


\end{itemize}

\subsection{Organization and Notation}

The rest of the paper is organized as follows: Section~\ref{system model} introduces the HISAC system and formulates the pattern design problem. Section~\ref{continuous-dis} presents the continuous-discrete transformation and reformulates the problem as a joint transmit–receive beamforming design. Section~\ref{solution} describes the proposed optimization methods, and Section~\ref{sec:Sims} provides numerical results. Finally, Section \ref{sec:Conclusions} concludes the paper.

Throughout the paper, bold lowercase and uppercase letters denote vectors and matrices, respectively. Notations include $\lambda_{\max}(\boldsymbol{A})$ for the largest eigenvalue, $\Vert\cdot\Vert_2$ and $\Vert\cdot\Vert_F$ for the $\ell_2$ and Frobenius norms, $(\cdot)^*$, $(\cdot)^T$, and $(\cdot)^H$ for complex conjugate, transpose, and Hermitian, respectively, and $\mathbf{E}(\cdot)$ for expectation. The identity matrix of size $N$ is denoted by $\boldsymbol{I}_N$.

\section{System Model and problem formulation}
\label{system model}

\subsection{HISAC Transceiver Architecture}

\begin{figure}[htbp]
\vspace{-0.5cm}
    \centering
    \includegraphics[width=.85\linewidth]{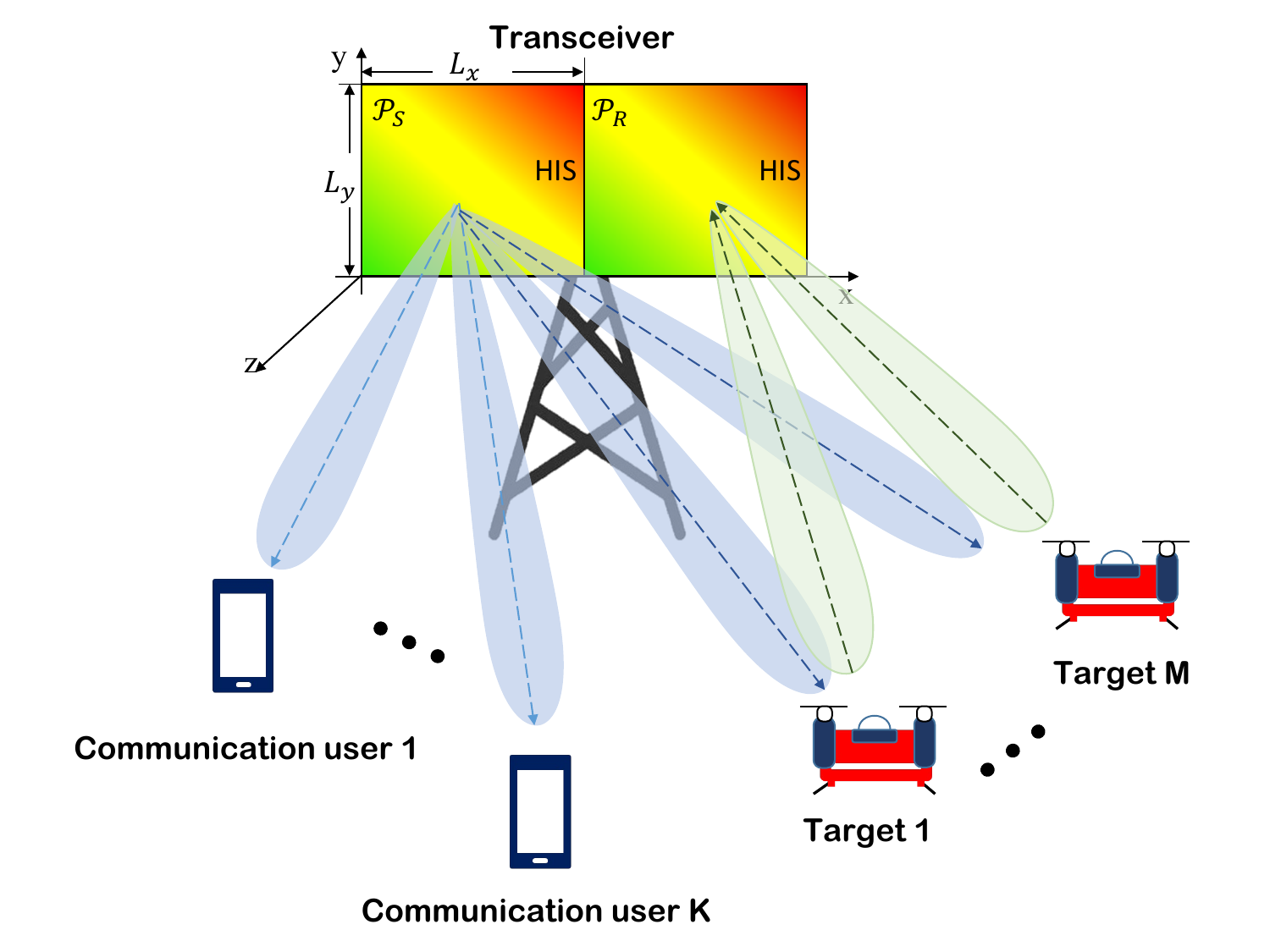}
    \caption{The illustration of the HIS-assisted integrated sensing and communication system.}
    \label{fig:hisac}
\end{figure}

\begin{figure}[htbp]
\vspace{-0.5cm}
    \centering
    \includegraphics[width=.85\linewidth]{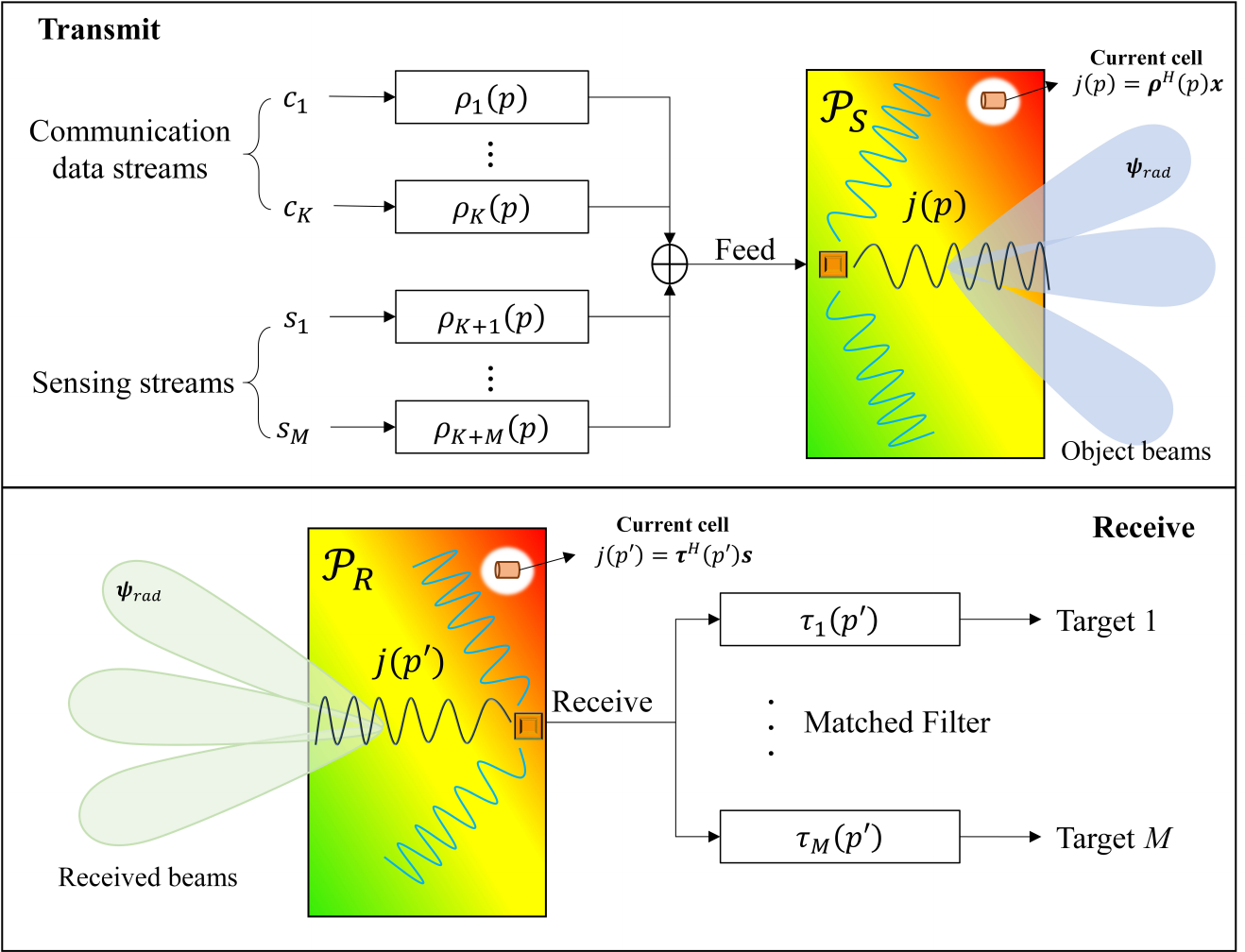}
    \caption{Illustration of the proposed HISAC transceiver structure.}
    \label{fig:his}
\end{figure}

As shown in Fig. \ref{fig:hisac}, we consider multi-user communication and multi-target sensing with the assistance of \acp{his}.
We introduce an \ac{hisac} system consisting of a transmit \ac{his} $\mathcal{P}_S$ and a receive \ac{his} $\mathcal{P}_R$. The aperture size for both \acp{his} is $A_T = L_x L_y$, where $L_x$ and $L_y$ are the lengths along $x$-axis and $y$-axis, respectively.  
Specifically, $\mathcal{P}_S$ sends signals to perform multi-target sensing and communicate to $K$ single-antenna users simultaneously, while $\mathcal{P}_R$ receives the radar echoes from $M$ targets. 
The structure of the \ac{hisac} transceiver operation is depicted in Fig. \ref{fig:his}. The proposed \ac{his} incorporates a single feed directly connected to the \ac{rf} distribution network within the surface. The traveling wave originates from the input port, propagates outward along the feed path, and excites the corresponding radiated wave \cite{karimipour2019shaping}. 
By carefully designing the surface current distribution $j(p)$, the radiation characteristics $\psi_{\text{rad}}(p)$ can be flexibly controlled, enabling the desired \ac{em} environment for \ac{isac} applications.
Motivated by this capability, we propose to modulate both the communication data and sensing streams into the surface current pattern of the \ac{his}. 
Specifically, the communication data streams and sensing streams are modulated into the transmit pattern at $\mathcal{P}_S$. Similarly, the echoed signals from $M$ targets are received by $\mathcal{P}_R$, and then demodulated by $M$ receive patterns. 

Let $\boldsymbol{x}=[\boldsymbol{c}^T,\boldsymbol{s}^T]^T$, where $\boldsymbol{c}=[c_1,..,c_K]^T\sim \mathcal{C N}\left(0, \boldsymbol{I}_K\right)$ and $\boldsymbol{s}=[s_1,..,s_M]^T$ are communication data streams and sensing streams, respectively. In particular, the sensing signal sequence $\boldsymbol{s}$ is assumed to have unit power and be uncorrelated with other streams, which is generated by random phase coding \cite{4}. Correspondingly, the transmit pattern at point $p$ of the surface is defined as $\boldsymbol{\rho}(p)=[\rho_1,...,\rho_{K+M}]^H$. 
The pattern generated by the communication data streams is given by $j_c(p)=\sum_{k=1}^K c_k \rho_k(p)$ \cite{zhang2023pattern}.
Similarly, the pattern generated by the sensing streams is given as $j_r(p)=\sum_{m=1}^M s_m \rho_{K+m}(p)$.
The combined pattern at $\mathcal{P}_S$ is given by
\begin{equation}
\label{j}
    \begin{aligned}
        \centering
        j(p)=\sum_{k=1}^K c_k \rho_k(p)+\sum_{m=1}^M s_m \rho_{K+m}(p)=\boldsymbol{\rho}^H(p)\boldsymbol{x}.
    \end{aligned}
\end{equation}
As shown in Fig.~\ref{fig:his}, by exciting each current cell at $\mathcal{P}_S$ with the modulated current $j(p)=\boldsymbol{\rho}^H(p)\boldsymbol{x}$, the \ac{isac} signal is embedded into the \ac{his} and subsequently radiated.

Similar to the transmit end, the received signal is filtered by a set of receive patterns $\boldsymbol{\tau}=[\tau_1,...,\tau_{M}]^H$ at $\mathcal{P}_R$ to extract the target information.  
Each receive current cell $j(p')=\boldsymbol{\tau}^H(p')\boldsymbol{s}$ is designed to match the specific target, enabling accurate separation and generating the receive beams.

To fully exploit the reconfigurability of the \ac{his} and maximize system performance, both $\boldsymbol{\rho}$ and $\boldsymbol{\tau}$ are jointly optimized to simultaneously improve sensing accuracy and increase communication capacity.
In the following, we consider \ac{sinr} as the performance metric of both communication and sensing, described as follows.

\subsection{Communications and Radar Performances}
\label{com_sinr}
Let $G_k(p)\triangleq G\left(p^{\prime}_k, p\right)$ be the Green's function from the source $p$ to the user $k$. Based on the principle of \ac{em} propagation theory, the electric field in the space domain is determined by $\boldsymbol{\rho}$ over surface $D$. The received signal at user $k$ can be calculated by \cite{Sanguinetti2022wavenumber}
\begin{equation}
\label{electric field}
    \begin{aligned}
        \centering
        e_k=&j \kappa Z_0 \int_D G_k(p) \rho_k(p)c_k \mathrm{d}p \\
        &+ \sum_{i=1,i\neq k}^{K+M} j \kappa Z_0 \int_D G_k(p) \rho_i(p)x_i \mathrm{d}p +v_c ,~k=1,...,K,
    \end{aligned}
\end{equation}
where $v_c$ is the \ac{awgn} with variance $\sigma_c^2$, $\kappa$ is the wavenumber, and $Z_0$ is the wave impedance in the free space. The communications \ac{sinr} of the $k$-th user, for $k=1,...,K$, is given by
\begin{equation}
\label{etac_con}
    \begin{aligned}
        \centering
        \eta_c(\boldsymbol{\rho}(p);k)=\frac{\mathbf{E}\left(\left|j \kappa Z_0 \int_D G_k(p) \rho_k(p)c_k \mathrm{d}p\right|^2\right)}{\mathbf{E}\left(\left|\sum_{i=1,i\neq k}^{K+M} j \kappa Z_0 \int_D G_k(p) \rho_i(p)x_i \mathrm{d}p\right|^2\right)+\sigma_c^2},
    \end{aligned}
\end{equation}

In our considered \ac{hisac} system, both communication data streams and radar streams are utilized to support multi-target sensing since they are known by the $\mathcal{P}_R$ from prior sensing steps.
In particular, we consider sensing SINR to design a focused HIS beampattern toward regions of interest identified in a previous sensing stage, thereby enabling more refined sensing tasks.
Following the standard full-duplex \ac{isac} assumption \cite{liu2022integrated,4,2,hua2023optimal,wen2023efficient}, \ac{si} after cancellation is assumed to be below the noise floor and is therefore omitted \cite{sabharwal2014inband,liu2022joint}.
Considering both transmit and receive patterns, the sensing \ac{sinr} is formulated in the following. 

Let $G_m(p)\triangleq G\left(p^{\prime}_m, p\right)$ and $G_m^*(p)\triangleq G\left(p,p^{\prime}_m\right)$ denote the Green’s function from source $p$ to $m$-th targets and from the $m$-th targets to source $p$, respectively. 
The $m$-th target's echo at $p^{\prime}$ is expressed as
\begin{equation}
\label{yp}
    \begin{aligned}
        \centering
        y_m(p')= \alpha_m G_m^*(p) e_m,~m=1,...,M,~p'\in \mathcal{P}_R,
    \end{aligned}
\end{equation}
where $e_m=j \kappa Z_0 \int_D G_m(p) \boldsymbol{\rho}(p)\boldsymbol{x} \mathrm{d}p$, and $\alpha_m$ is the backscattering intensity of the $m$-th target. For ease of presentation and without loss of generality, each target's backscattering intensity is considered the same and ignored in the following. 

Similarly, the received field at $\mathcal{P}_R$ is determined by $\boldsymbol{\tau}$ over surface $D$. 
Then, the received signal with respect to the $l$-th target is given by
\begin{equation}
\label{yl}
    \begin{aligned}
        \centering
        y_l=&\sum_{m=1}^{M}\int_D \tau_l\left(p^{\prime}\right) y_m\left(p^{\prime}\right) \mathrm{d} p^{\prime} + \int_D \tau_l(p^{\prime}) v_r\left(p^{\prime}\right) \mathrm{d} p^{\prime},
    \end{aligned}
\end{equation}
where ${v}_r(p'),~p'\in D$ is the \ac{awgn} at $\mathcal{P}_R$ with variance $\sigma_r^2$.
Therefore, for $l=1,...,M$, the sensing \ac{sinr} is given by 
\begin{equation}
\label{etar_con}
    \begin{aligned}
        \centering    \eta_r(\boldsymbol{\rho},\boldsymbol{\tau};l)=\frac{\mathbf{E}\left(\left|e_l \int_D \tau_l\left(p^{\prime}\right) G_l^*(p) \mathrm{d} p^{\prime}\right|^2\right)}{\mathbf{E}\left(\left|\sum_{m=1,m\neq l}^{M}e_m \int_D \tau_l\left(p^{\prime}\right) G_m^*(p) \mathrm{d} p^{\prime}\right|^2\right)+\sigma_r^2}.
    \end{aligned}
\end{equation}

\subsection{Problem Formulation in Continuous Domain}

From (\ref{etac_con}) and (\ref{etar_con}), both the sensing and communication \acp{sinr} rely on $\boldsymbol{\rho}$. Meanwhile, the sensing \ac{sinr} is affected by $\boldsymbol{\tau}$. 
Here, we aim to strike a performance trade-off between sensing and communication in the proposed \ac{hisac} system. Specifically, we propose to maximize the minimal sensing \ac{sinr} among multiple targets while guaranteeing the multi-user communication \ac{sinr} requirement. The resulting problem w.r.t. the continuous pattern design is formulated as
\begin{subequations}
\label{P00}
    \begin{align}
        \centering
        \max_{\boldsymbol{\rho},\boldsymbol{\tau}} \ \min_{l=1,...,M} &\ \eta_r(\boldsymbol{\rho},\boldsymbol{\tau};l), \label{eta_r_p0}\\
    {\rm s.t.} &\ \eta_c(\boldsymbol{\rho};k)\geq \Gamma_c, ~k=1,...,K \label{eta_c_p0},\\
    &\ \int_D \Vert\boldsymbol{\rho}\Vert_2^2 \mathrm{d}p \leq P_T \label{power_w0},\\
    &\ \int_D |{\tau}_l |^2 \mathrm{d}p' =1,~l=1,...,M \label{power_q0}.
    \end{align}
\end{subequations}

The primary goal of (\ref{P00}) is to balance the sensing \ac{sinr} among radar targets, and (\ref{eta_c_p0}) is the communication constraint for multiple users, where $\Gamma_c$ represents the minimal required communication \ac{sinr}. Concerning power constraints, (\ref{power_w0}) and (\ref{power_q0}) represent the power budget for the transmit pattern at $\mathcal{P}_S$ and the unit-power constraint for the receive pattern at $\mathcal{P}_R$, respectively. 
Generally, the receive beamformer's power in the direction of interest is maintained at $1$ to ensure a distortionless response for the signal coming from that specific direction, where receive beamforming vectors satisfy $|{\tau}_l|^2 \mathrm{d}p' =1$.

Due to the presence of infinite-dimensional integrals in both the objective and constraints of problem (\ref{P00}), classical beamforming techniques designed for discrete-array-based \ac{isac} systems are not directly applicable.
Moreover, as illustrated in Fig.~\ref{fig:his}, individually designing each current cell on the HIS surface is impractical from a hardware perspective, motivating the need for a more feasible alternative.
To address the challenges posed by the infinite-dimensional functions in problem (\ref{P00}), we develop a continuous-discrete transformation inspired by the Fourier transform and propose a tractable hardware implementation of the HIS, as detailed in the following section.

\section{Continuous-Discrete Transformation}
\label{continuous-dis}

In this section, we formulate a comprehensive continuous-discrete transformation for $\boldsymbol{\rho}$, which is then applied to derive the expressions for communication and sensing \acp{sinr}.
Based on this transformation, we approximate the continuous pattern design in (\ref{P00}) into a joint transmit-receive beamforming problem in the discrete domain.

\subsection{Comprehensive Continuous-Discrete Transformation}
\label{ccdt}

Due to the continuity of pattern at \ac{his}, we first transform patterns into discrete domains. For an integrable $\boldsymbol{\rho}$ defined over surface $D$, $\boldsymbol{\rho}(p)$ can be equivalently converted from the space domain into the wavenumber domain as \cite{Sanguinetti2022wavenumber}
\begin{equation}
\label{rho_Fourier}
    \begin{aligned}
        \centering
        \rho_i(p)=\sum_n^{\infty} w_{i, n} \Psi_n(p),~i=1,...,K+M,
    \end{aligned}
\end{equation}
where $w_{i,n}$ is the coefficient of the $i$-th pattern projected to the $n$-th Fourier transform function $\Psi_n(p)$. 
By substituting (\ref{rho_Fourier}) into (\ref{electric field}), the electric field is represented by
\begin{equation}
    \label{e_Fourier}
        \begin{aligned}
            \centering
            e\left(p^{\prime}\right)=
            j \kappa Z_0 \sum_{i=1}^{K+M}\sum_n^{\infty} x_i w_{i, n} f_{2 D}^n\left(G_{p^{\prime}}\right) +v_c,
        \end{aligned}
    \end{equation}
where $f_{2 D}^n\left(G_{p^{\prime}}\right)\triangleq \int_D G\left(p^{\prime}, p\right) \Psi_n(p) \mathrm{d} p$ is the Green's function Fourier transform over surface $D$.

To overcome the integral functions in the pattern design, we provide a closed-form expression for $f_{2 D}^n\left(G_{p^{\prime}}\right)$ in the following.

The expression of the scalar Green's function in the far-field is given by
\begin{equation}
\label{GF}
    \begin{aligned}
        \centering
        G(\textbf{r},\textbf{s}) = \frac{e^{j\kappa \|\textbf{r}-\textbf{s}\|}}{4\pi \|\textbf{r}-\textbf{s}\|},
    \end{aligned}
\end{equation}
where $\textbf{r}$ and $\textbf{s}$ are positions of the point target and source, respectively. Let $\textbf{r}=(r, \theta, \psi)$ denote the position of ${p^{\prime}}$ in the Spherical coordinate, and $\textbf{s}=(s_x, s_y)$ be a source point at the surface in the Cartesian coordinate. The Euclidean distance in the Cartesian coordinate system is calculated as
\begin{equation}
\label{distance}
    \begin{aligned}
        \centering
        \| \textbf{r}- \textbf{s}\| 
        \approx r-\sin{\theta}\left(s_x\cos{\psi}+s_y\sin{\psi}\right).
    \end{aligned}
\end{equation}
By substituting (\ref{distance}) into (\ref{GF}), the Green's function on the far field is approximated as
\begin{equation}
\label{GF_far}
    \begin{aligned}
        \centering
        G_{p^{\prime}} = \frac{e^{j\kappa r}}{4\pi r}e^{-j\kappa s_x \sin{\theta}\cos{\psi}}e^{-j\kappa s_y \sin{\theta}\sin{\psi}}.
    \end{aligned}
\end{equation}
Combining (\ref{GF_far}) and (\ref{e_Fourier}), the Green's function Fourier transform $f_{2 D}^n\left(G_{p^{\prime}}\right)$, is given by
\begin{equation}
\label{f_2D_Gm_far}
    \begin{aligned}
        \centering
        f_{2 D}^n\left(G_{p^{\prime}}\right)=
        \int_D \frac{e^{j\kappa r}}{4\pi r}e^{-j\kappa s_x \sin{\theta}\cos{\psi}}e^{-j\kappa s_y \sin{\theta}\sin{\psi}} \Psi_n(p) \mathrm{d} p,
    \end{aligned}
\end{equation}
where we decompose the $n$-th order $n=(n_x,n_y)$ along the $x$-axis and $y$-axis, the Fourier transform function $\Psi_n(p)$ is rewritten as
\begin{equation}
\label{Psi_n}
    \begin{aligned}
        \centering
        \Psi_n(p)=\frac{1}{\sqrt{A_T}}e^{-j2\pi \left(\frac{n_x}{L_x}\left(s_x-\frac{L_x}{2}\right)\right)}e^{-j2\pi \left(\frac{n_y}{L_y}\left(s_y-\frac{L_y}{2}\right)\right)}.
    \end{aligned}
\end{equation}

Let $\kappa_x^n = \kappa \left(\sin{\theta}\cos{\psi}+\lambda\frac{n_x}{L_x}\right)$ and $\kappa_y^n = \kappa \left(\sin{\theta}\sin{\psi}+\lambda\frac{n_y}{L_y}\right)$ denote the wavenumbers along the $x$-axis and $y$-axis, respectively. 
Combining (\ref{f_2D_Gm_far}) with (\ref{Psi_n}), the closed-form Green's function Fourier transform for arbitrary position $(r, \theta,\psi)$ can be calculated by (\ref{f_2D_2}). 


\begin{equation}
\label{f_2D_2}
    \begin{aligned}
        \centering
        f_{2 D}^n &\left(G_{p^{\prime}}\right)=\frac{e^{j\kappa r}}{4\pi r\sqrt{A_T}} e^{j(n_x+n_y) \pi}\\
        &\int_{L_x}e^{-j\kappa_x^n s_x}\mathrm{d}s_x 
        \int_{L_y}e^{-j\kappa_y^n s_y}\mathrm{d}s_y \\
        =&\frac{e^{j\kappa r}e^{j(n_x+n_y) \pi}}{4\pi r\sqrt{A_T}} \left[\frac{e^{-j\kappa_x^n s_x}}{-j\kappa_x^n}\right]_{-\frac{L_x}{2}}^{\frac{L_x}{2}}\left[\frac{e^{-j\kappa_y^n s_y}}{-j\kappa_y^n}\right]_{-\frac{L_y}{2}}^{\frac{L_y}{2}} \\
        =& \frac{e^{j\kappa r}\sqrt{A_T}}{4\pi r} e^{j(n_x+n_y) \pi} \mathrm{sinc}{\left(\kappa_x^n \frac{L_x}{2}\right)}\mathrm{sinc}{\left(\kappa_y^n \frac{L_y}{2}\right)}.
    \end{aligned}
\end{equation}
It's obvious that $f_{2 D}^n\left(G_{p^{\prime}}\right)$ is determined by the $\kappa_x^n$ and $\kappa_y^n$. Taking the $x$-axis as example, when the $x$-axis wavenumber satisfies
\begin{equation}
\label{first_zero}
    \begin{aligned}
        \centering
        \kappa \left(\sin{\theta}\cos{\psi}+\lambda\frac{n_x}{L_x}\right) \cdot \frac{L_x}{2} \geq n\pi,
    \end{aligned}
\end{equation}
the corresponding $\mathrm{sinc}$ term approaches its first zero, indicating that higher-order terms contribute negligibly to the total response and can be safely truncated. Consequently, the truncation order $(n_x,n_y)$ should satisfy $\{|n_x| \geq \frac{L_x}{\lambda};|n_y| \geq \frac{L_y}{\lambda}\}$, and the minimum total truncation order is bounded by $N\geq \left(2\frac{L_x}{\lambda}+1\right)\left(2\frac{L_y}{\lambda}+1\right)$.
It is worth clarifying that the spatial DoFs of the aperture are fundamentally determined by its electrical size \cite{dardari2020communicating} and do not depend on the truncation order itself. 
In particular, the maximum number of independent spatial DoFs supported by the continuous-aperture is bounded by $\frac{\pi A_T}{\lambda^2}$.
The condition in (\ref{first_zero}) specifies a sufficient truncation order such that the retained Fourier modes fully capture the intrinsic DoFs supported by the continuous-aperture. When this condition is satisfied, higher-order terms contribute negligibly, and increasing the truncation order beyond this point does not yield additional DoFs.

\subsection{Representation of Communication and sensing SINRs}
\label{rcsinr}

Based on the previous transformation, the communication \ac{sinr} can be described as follows.
Combining (\ref{f_2D_2}) with (\ref{electric field}), the received signal at user $k$ is calculated as
\begin{equation}
    \label{ek}
        \begin{aligned}
            \centering
            \hat{e}_k=j \kappa Z_0\sum_{i=1}^{K+M}\sum_n^{\infty} x_i w_{i, n} f_{2 D}^n\left(G_k\right)+v_c,~k=1,...,K.
        \end{aligned}
    \end{equation}

According to \cite{demir2022channel,Sanguinetti2022wavenumber}, the channel gain $\sum_n^{\infty}f_{2 D}^n\left(G_k\right)$ within the band of $[- \kappa, \kappa]$ dominates the radiation power over the Fourier wavenumber domain. According to (\ref{first_zero}), the Fourier series $f_{2 D}^n\left(G_k\right)$ can be approximated up to the $N$-th order, comprising vector
\begin{equation}
\label{f_2d_vec}
    \begin{aligned}
        \centering
        \boldsymbol{f}_k=[f_{2 D}^1\left(G_k\right),...,f_{2 D}^N\left(G_k\right)]^H, ~k=1,...,K.
    \end{aligned}
\end{equation}
For convenience, we also collect the coefficients of the $i$-th pattern into a vector (beamformer) 
\begin{equation}
\label{w_i}
    \begin{aligned}
        \centering
        \boldsymbol{w}_i = [w_{i,1},...,w_{i,N}]^H, ~i=1,...,K+M.
    \end{aligned}
\end{equation}
Then, the beamforming matrix $\boldsymbol{W}$ of the whole pattern is expressed by $\boldsymbol{W}=[\boldsymbol{W}_c,\boldsymbol{W}_{r}]$, where $\boldsymbol{W}_c\triangleq [\boldsymbol{w}_1,...,\boldsymbol{w}_{K}]$ and $\boldsymbol{W}_r\triangleq [\boldsymbol{w}_{K+1},...,\boldsymbol{w}_{K+M}]$ are the beamforming matrices for communication data streams and sensing streams, respectively.
By combining equations (\ref{ek}), (\ref{f_2d_vec}), and (\ref{w_i}), we approximate the received signal at the $k$-th user as
\begin{equation}
\label{ek_matrix}
    \begin{aligned}
        \centering
        \mathcal{E}_k=& j \kappa Z_0 \boldsymbol{f}_k^H \boldsymbol{w}_k c_k + \underbrace{\sum_{i=1,i\neq k}^{K} j \kappa Z_0 \boldsymbol{f}_k^H \boldsymbol{w}_i c_i}_\mathrm{Multi-user~Interference} \\
        &+ \underbrace{j \kappa Z_0 \boldsymbol{f}_k^H \boldsymbol{W}_r\boldsymbol{s}}_\mathrm{Sensing~Interference} + v_c,~k=1,...,K.
    \end{aligned}
\end{equation}
{Therefore, combining (\ref{etac_con}) with (\ref{ek_matrix}), the \ac{sinr} at the $k$-th user is expressed by (\ref{eta_c}).}
\begin{figure*}[htb]
    \begin{equation}
    \label{eta_c}
    \begin{aligned}
        \centering
        \eta_c(\boldsymbol{W};k) = \frac{\boldsymbol{f}_k^H \boldsymbol{w}_k\boldsymbol{w}^H_k \boldsymbol{f}_k}{\boldsymbol{f}_k^H \boldsymbol{W}\boldsymbol{W}^H \boldsymbol{f}_k-\boldsymbol{f}_k^H \boldsymbol{w}_k\boldsymbol{w}^H_k \boldsymbol{f}_k+\frac{\sigma_c^2}{\kappa^2 Z^2_0}}, ~k=1,...,K.
    \end{aligned}
\end{equation}
\begin{equation}
        \label{eta_r}
            \begin{aligned}
                \centering
                \eta_r(\boldsymbol{W},\boldsymbol{q}_l;l)=\frac{ \boldsymbol{q}_l^H\boldsymbol{G}_l\boldsymbol{WW}^H\boldsymbol{G}^H_l\boldsymbol{q}_l}{\boldsymbol{q}_l^H \left(\sum_{m=1}^{M}\boldsymbol{G}_m\boldsymbol{WW}^H\boldsymbol{G}^H_m -\boldsymbol{G}_l\boldsymbol{WW}^H\boldsymbol{G}^H_l \right)\boldsymbol{q}_l+\frac{\sigma_r^2}{\kappa^2 Z^2_0}},~l=1,...,M.
            \end{aligned}
    \end{equation}
    \hrulefill
\end{figure*}

Similar to the transmit end, the receive pattern in the continuous space domain is converted into the discrete Fourier domain by $\tau_l(p')=\sum_n^{\infty} q_{l, n} \Phi_n(p')$,
where $q_{l,n}$ is the coefficient of the $l$-th receive pattern projected to $\Phi_n(p')$.
We confine the Fourier series $g_{2 D}^n\left(G^*_m\right)$ over surface $\mathcal{P}_R$ to the $N$-th order and comprise vector $\boldsymbol{g}_{m}^*=[g_{2 D}^1\left(G_m^*\right),...,g_{2 D}^N\left(G_m^*\right)]^H$. 
The coefficients of the $l$-th pattern are collected into a vector $\boldsymbol{q}_l=[q_{l,1},...,q_{l,N}]^H$.
Combining (\ref{f_2D_2}) with (\ref{yl}), the received echo w.r.t. the $l$-th target is approximated as
\begin{equation}
\label{yl_vec}
    \begin{aligned}
        \centering
        \mathcal{Y}_{l} = &\underbrace{j \kappa Z_0 \boldsymbol{q}_l^H \boldsymbol{G}_l \boldsymbol{W} \boldsymbol{x}}_\mathrm{Target~echo} + \underbrace{\sum_{m=1,m\neq l}^M j \kappa Z_0 \boldsymbol{q}_l^H \boldsymbol{G}_m \boldsymbol{W} \boldsymbol{x}}_\mathrm{Interference~echo}\\
        &+\underbrace{\boldsymbol{q}_l^H\boldsymbol{v}_r}_\mathrm{Beamformer~Noise},~l=1,...,M,
    \end{aligned}
\end{equation}
where $\boldsymbol{G}_m=\boldsymbol{g}_{m} \boldsymbol{g}_m^H$, and $\boldsymbol{v}_r$ is the Fourier transform of ${v}_r$ over surface $D$ via the truncated oreder $N$.  

\begin{remark}
\label{remark}
The discrete-domain receive noise vector $\boldsymbol{v}_r$ obtained via the truncated Fourier representation follows an i.i.d. circularly symmetric complex Gaussian distribution with covariance $\boldsymbol{v}_r\sim \mathcal{C N}\left(0, \sigma_r^2\boldsymbol{I}_N\right)$.
\end{remark}
\begin{IEEEproof}
See Appendix. \ref{remark1}
\end{IEEEproof}

Accordingly, the receive beamformer $\boldsymbol{q}_l$ of the \ac{hisac} system ensures that the signal of interest is received without amplification and attenuation, allowing for accurate reception and signal processing.
With unit-power beamformer $\boldsymbol{q}_l$ at $\mathcal{P}_R$, the corresponding sensing \ac{sinr} w.r.t. the $l$-th target in (\ref{etar_con}) is then expressed as (\ref{eta_r}).
\begin{figure}[htbp]
    \centering
    \includegraphics[width=.95\linewidth]{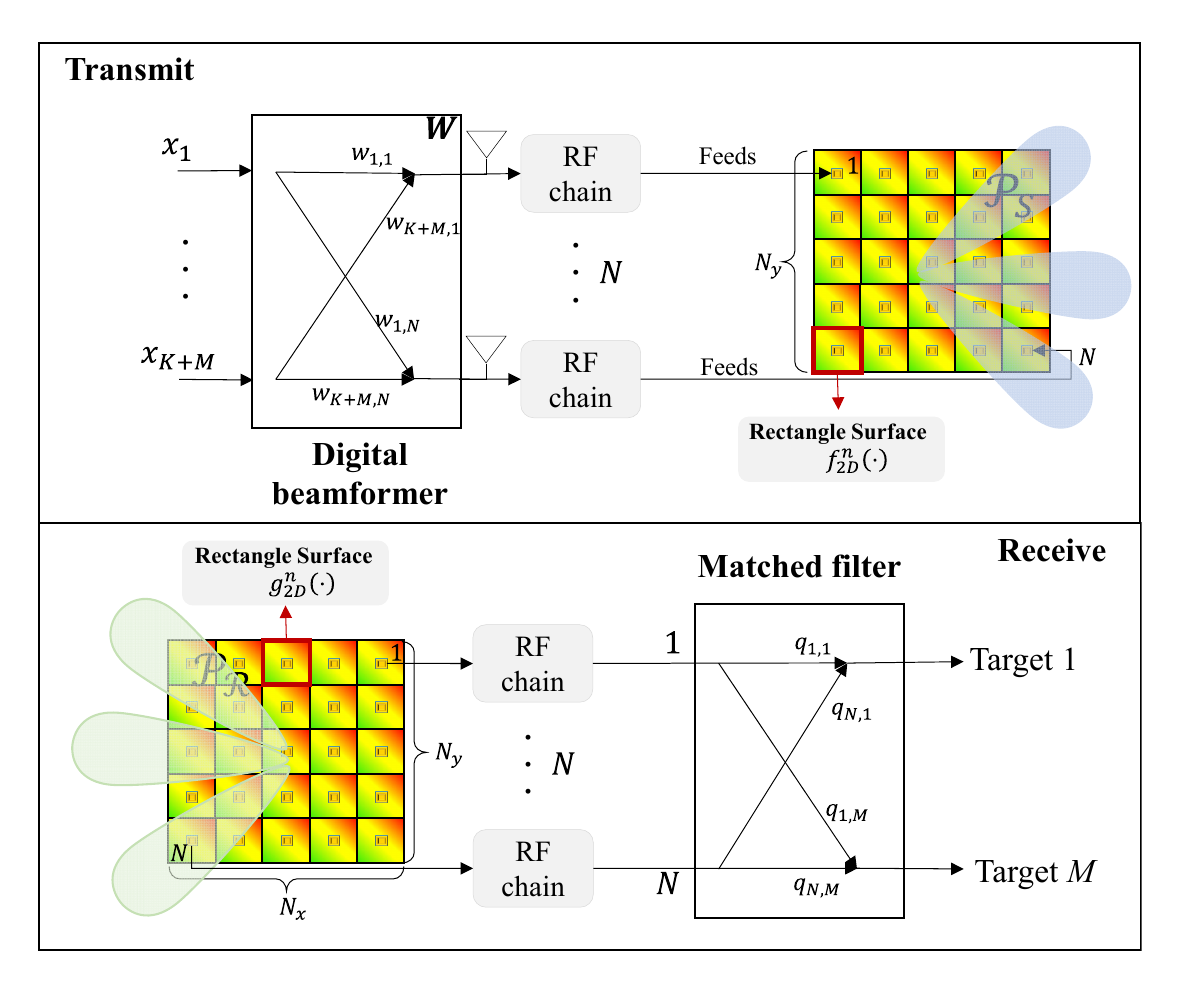}
    \caption{The fully-digital HIS architecture.}
    \label{fig:transmit_hardware}
\end{figure}

\subsection{Antenna Architecture}
According to the continuous-discrete transformation, the HIS beam pattern can be reconstructed by summing $N$ beams with specific wavenumbers. Consequently, the continuous beam pattern of the HIS can be effectively approximated using a discrete set of $N$ beams.
To achieve the continuous-discrete transformation of the \ac{his} and optimize its pattern, we consider the fully-digital architecture, as shown in Fig.~\ref{fig:transmit_hardware}. In the following subsections, we briefly introduce the antenna architecture and discuss its corresponding beamformer models.

\subsubsection{Fully-Digital Architecture}
In the fully-digital antenna architecture, each beam’s feed is connected to a dedicated \ac{rf} chain, as shown in Fig.~\ref{fig:transmit_hardware}. 
This configuration provides the \ac{hisac} system with maximum flexibility in signal processing, allowing independent control over the patterns of each beam.
Within this framework, the transmit \ac{his} beamforming design corresponds directly to the Fourier transform coefficient matrix $\boldsymbol{W}$ in (\ref{eta_c}) and (\ref{eta_r}). Similarly, the receive \ac{his} beamforming is implemented through simultaneous reception, which aligns with the coefficient vectors $\boldsymbol{q}_l$ in (\ref{eta_r}). Consequently, leveraging the principle of continuous-discrete transformation, \ac{his}'s beam can be turned by designing the \(\boldsymbol{W}\) matrix within the fully-digital architecture.

To implement the continuous HIS in hardware, we construct the surface using $N=N_x\times N_y$ square-shaped radiators, densely tiled to form the full aperture. Here, $N_x$ and $N_y$ denote the number of truncated Fourier orders along the $x$- and $y$-axes, respectively. Each radiator occupied an area of $A_T/(N_x-1)(N_y-1)$ serves as a sub-element that supports programmable radiation or reception characteristics.
On the transmit side, the radiation function of each unit is operated by the basis function $f^n_{2D}(\cdot)$, while on the receive side, the directional response is represented by $g^n_{2D}(\cdot)$.
The overall \ac{em} behavior of the HIS is then synthesized through the weighted superposition of these 
$N$ radiators.

Unlike discrete arrays that use dipole-like isotropic radiators, \ac{his} employs a continuous surface composed of densely packed rectangular patches to excite surface currents \cite{pizzo2022fourier,zhang2023pattern}, enabling enhanced beamforming control.
Note that the number of \ac{rf} chains in our fully-digital architecture is set to be equal to the truncation order $N$, ensuring one-to-one mapping between digital control and spatial basis functions.
In future work, the proposed design may be extended to hybrid architectures with fewer RF chains than $N$, where spatial multiplexing can be achieved by phase shifters.


\subsection{Problem Formulation in Discrete Domain}

Based on the above antenna architecture, we consider beamforming matrices $\boldsymbol{W}$ and $\boldsymbol{Q}$ as optimization variables.
From (\ref{eta_c}) and (\ref{eta_r}), both the sensing and communication \acp{sinr} are converted into the discrete domain and determined by transmit beamformers in $\boldsymbol{W}$ and receive beamformers collected in $\boldsymbol{Q}=[\boldsymbol{q}_1,...,\boldsymbol{q}_M]$. 
The original problem in (\ref{P00}) is converted from the continuous domain into the discrete domain. Based on the transformation in Section~\ref{ccdt} and truncation in Section~\ref{rcsinr}, the resulting approximated problem w.r.t. the joint transmit-receive beamforming can be formulated as
\begin{subequations}
\label{P0}
    \begin{align}
        \centering
        \max_{\boldsymbol{W},\boldsymbol{Q}} \ \min_{l=1,...,M} &\ \eta_r(\boldsymbol{W},\boldsymbol{q}_l;l), \label{eta_r_p}\\
    {\rm s.t.} &\ \eta_c(\boldsymbol{W};k)\geq \Gamma_c, ~k=1,...,K \label{eta_c_p},\\
    &\ \Vert \boldsymbol{W} \Vert_F^2 \leq P_T \label{power_w},\\
    &\ \Vert \boldsymbol{q}_l \Vert_2^2 = 1,~l=1,...,M \label{power_q},
    \end{align}
\end{subequations}
where $\eta_r(\boldsymbol{W},\boldsymbol{q}_l;l)$ and $\eta_c(\boldsymbol{W};k)$ are defined by (\ref{eta_r}) and (\ref{eta_c}), respectively. 
In \cite{zhang2023pattern}, the authors provide a methodology for computing the physical radiation power for a given surface configured by multiple patterns.
The total transmit power of transmit patterns can be upper-bounded by \cite{zhang2023pattern}
\begin{equation}
\label{power constraint}
    \begin{aligned}
        \centering
        \int_D \Vert\boldsymbol{\rho}(p)\Vert_2^2 \mathrm{d}p = \Vert \boldsymbol{W} \Vert_F^2\leq P_T.
    \end{aligned}
\end{equation}
In (\ref{power_q}), the unit-power requirement for receive beamformers at $\mathcal{P}_R$ is indicated.

Due to the non-convex max-min objective function and coupled optimized variables in objective and constraints, (\ref{P0}) is non-convex and challenging to solve. To circumvent the difficulties, we develop an efficient \ac{ao} algorithm to address the joint transmit-receive beamforming in the \ac{hisac} system.

\section{Joint Transmit-Receive Beamforming in HISAC}
\label{solution}

{In this section, we develop an \ac{ao} algorithm to address the joint transmit-receive beamforming in our proposed \ac{hisac} system under fully-digital architecture.}
We begin with introducing the transmit \ac{his} beamforming in Section~\ref{sec:transmit-design}, which we then utilize to design the receive beamforming with fixed transmit beamforming in Section~\ref{sec: receive design}.
In addition, we analyze the complexity and convergence of the proposed algorithm in Section~\ref{sec:conplexity}.

\subsection{HIS Transmit Beamforming with Fixed Receive Beamforming}
\label{sec:transmit-design}

In this section, we optimize the transmit beamforming matrix $\boldsymbol{W}$ with fixed receive beamforming matrix $\boldsymbol{Q}$. In particular, we propose a \ac{sdr}-based algorithm to obtain the optimal solution of (\ref{P0}) iteratively. Then, we develop an \ac{abs} method to efficiently accelerate the convergence speed.

\subsubsection{SDR-Based Algorithm}
\label{sdr-based}

For the convenience of analysis, we rewrite the \ac{sinr} of the sensing and communication in terms of the covariance matrix $\boldsymbol{R}$, represented by
\begin{equation}
    \label{eq_rc}
    \begin{aligned}
    \centering
        &\eta_c(\boldsymbol{R};k) = \frac{\boldsymbol{f}_k^H \boldsymbol{R}_k \boldsymbol{f}_k}{\boldsymbol{f}_k^H \left(\boldsymbol{R}-\boldsymbol{R}_k\right)\boldsymbol{f}_k+\frac{\sigma_c^2}{\kappa^2 Z^2_0}},~k=1,...,K,\\
        &\eta_r(\boldsymbol{R};l)=\\
        &\frac{ \boldsymbol{q}_l^{H}\boldsymbol{G}_l\boldsymbol{R}\boldsymbol{G}^H_l\boldsymbol{q}_l}{\boldsymbol{q}_l^{H}\left(\sum_{m=1}^{M}\boldsymbol{G}_m\boldsymbol{R}\boldsymbol{G}^H_m -\boldsymbol{G}_l\boldsymbol{R}\boldsymbol{G}^H_l \right)\boldsymbol{q}_l+\frac{\sigma_r^2}{\kappa^2 Z^2_0}},\\
        &~~~~~~~~~~~~~~~~~~~~~~~~~~~~~~~~~~~~~~~~~~~~~~l=1,...,M,
    \end{aligned}
\end{equation}
where $\boldsymbol{R}_k=\boldsymbol{w}_k\boldsymbol{w}_k^H$ and $\boldsymbol{R}=\boldsymbol{W}\boldsymbol{W}^H$. With the given receive beamformer $\boldsymbol{Q}=\left[\boldsymbol{q}_1,...,\boldsymbol{q}_M\right]$, we propose to maximize the minimal sensing \ac{sinr} by optimizing the total covariance matrix $\boldsymbol{R}$ and its sub-element  $\boldsymbol{R}_k$. 

 

To deal with it, we first omit the rank-one constraints and transform (\ref{P0}) into a series of feasibility-checking \cite{zhang2010relationship} sub-problems by \ac{sdr} as
\begin{subequations}
\label{st3_}
    \begin{align}
    \centering
    \max_{\boldsymbol{R},\boldsymbol{R}_1,...,\boldsymbol{R}_{K},\Gamma_r} \quad  &\Gamma_r,\\
    {\rm s.t.} \ & \eta_r(\boldsymbol{R};l)\ge\Gamma_r,~l=1,...,M,\label{st3_er}\\
    &\eta_c(\boldsymbol{R};k)\ge\Gamma_c,~k=1,...,K,\label{st3_ec}\\
    & \boldsymbol{R} \succeq 0,\\
    & \boldsymbol{R}_k \succeq 0,~ k=1,...,K,\\
    & \boldsymbol{R}-\sum_{k=1}^{K}\boldsymbol{R}_k \succeq 0,\\
    & {\rm tr}(\boldsymbol{R})\leq P_T,
    \end{align}
\end{subequations}
where $\Gamma_r$ is an auxiliary variable to replace the minimal sensing \ac{sinr} in the objective function. 

Due to the fractional terms in constraints (\ref{st3_er}) and (\ref{st3_ec}), the above problem is still non-convex.
Let $\boldsymbol{U}_{l,m} = \boldsymbol{G}_m^H\boldsymbol{q}_l\boldsymbol{q}_l^{H}\boldsymbol{G}_m$, $\boldsymbol{F}_k = \boldsymbol{f}_k\boldsymbol{f}_k^H$.
By fixing $\Gamma_r$ and rearranging (\ref{st3_er}) and (\ref{st3_ec}) as
\begin{equation}
    \begin{aligned}
        \centering   (1+\Gamma_r^{-1}){\rm tr}\left(\boldsymbol{U}_{l,l}\boldsymbol{R}\right)-\sum_{m=1}^{M}{\rm tr}\left(\boldsymbol{U}_{l,m}\boldsymbol{R}\right) \geq \frac{\sigma_r^2}{\kappa^2 Z^2_0} \label{gr1},
    \end{aligned}
\end{equation}
and
\begin{equation}
    \begin{aligned}
        \centering        (1+\Gamma_c^{-1}){\rm tr}\left(\boldsymbol{F}_{k}\boldsymbol{R}_k\right)-{\rm tr}\left(\boldsymbol{F}_{k}\boldsymbol{R}\right) \geq \frac{\sigma_c^2}{\kappa^2 Z^2_0},
    \end{aligned}
\end{equation}
respectively.
(\ref{st3_}) becomes a convex \ac{sdp} feasibility-checking problem and can be solved by off-the-shelf toolbox such as CVX \cite{toh1999sdpt3}. 
By providing a potential range for $\Gamma_r$ that contains the optimal $\Gamma_r^*$, (\ref{st3_}) can be solved by checking its feasibility with $\Gamma_r$ being chosen in a bisection manner. {Specifically, we perform a bisection search over the interval $\left[\Gamma_{r,start},\Gamma_{r,end}\right]$}, as shown in Algorithm \ref{al0}.
\begin{algorithm}
    \caption{Bisection Search with SDR}
    \renewcommand{\algorithmicrequire}{\textbf{Input:}}
    \renewcommand{\algorithmicensure}{\textbf{Output:}}
    \label{al0}
    \begin{algorithmic}[1]
        \REQUIRE \leavevmode 
        $\Gamma_c$.
        \STATE Determine $\Gamma_{r,start}=0$, $\Gamma_{r,end}=\frac{\kappa^2 Z_0^2 P_T A_T}{\sigma_r^2}$.
        \REPEAT
        \STATE Update $\Gamma_r \leftarrow \frac{1}{2}\left(\Gamma_{r,strat}+\Gamma_{r,end}\right)$.
        \STATE Solve (\ref{st3_}) by checking its feasibility with $\Gamma_r$. If (\ref{st3_}) is feasible, $\Gamma_{r,start} \leftarrow \Gamma_r$; otherwise $\Gamma_{r,end} \leftarrow \Gamma_r$.
        \UNTIL $\Gamma_{r,end}-\Gamma_{r,start}\leq \epsilon_1$
        \STATE Update the minimal sensing SINR as $\Gamma_{r,start}$.
        \STATE Calculate $\boldsymbol{R},\boldsymbol{R}_1,...,\boldsymbol{R}_K$ by solving (\ref{st3_}) with $\Gamma_{r,start}$.
        \ENSURE \leavevmode 
        $\boldsymbol{R},\boldsymbol{R}_1,...,\boldsymbol{R}_K$.\\
    \end{algorithmic}
\end{algorithm}

Let $\boldsymbol{\hat{R}}_1,...,\boldsymbol{\hat{R}}_{K},\boldsymbol{\hat{R}}$ be the optimal solution of (\ref{st3_}), each entry of the communication beamforming matrix $\boldsymbol{\hat{W}}_c$ is calculated by 
    \begin{equation}
    \label{W_c}
    \begin{aligned}
        \centering
        \boldsymbol{\hat{w}}_k = \left(\boldsymbol{f}_k^H \boldsymbol{\hat{R}}_k \boldsymbol{f}_k\right)^{-\frac{1}{2}}\boldsymbol{\hat{R}}_k\boldsymbol{f}_k,~k=1,...,K,
    \end{aligned}
\end{equation}
    and the beamforming matrix $\boldsymbol{\hat{W}}_r$ for multi-target sensing is choosing by 
    \begin{equation}
    \label{W_r}
        \begin{aligned}
            \centering
            \boldsymbol{\hat{W}}_r\boldsymbol{\hat{W}}_r^H = \boldsymbol{\hat{R}}-\sum_{k=1}^{K}\boldsymbol{\hat{R}}_k.
        \end{aligned}
    \end{equation}
    The optimal solution $\boldsymbol{\bar{R}}_1,...,\boldsymbol{\bar{R}}_{K}$ for (\ref{P0}) is reconstructed as $\boldsymbol{\bar{R}}_{k}=\boldsymbol{\hat{w}}_k\boldsymbol{\hat{w}}_k^H$, for $k=1,...,K$, which is exactly rank-one and optimal.

It is noteworthy that, for different \ac{hisac} system setups, the dynamic searching range for potential values of feasible $\Gamma_r$ changes rapidly, leading to high overhead. To overcome this challenge, we propose an \ac{abs} method which is detailed in the following, which provides a refined/shrunk searching range for $\Gamma_r$ to accelerate the convergence of {Algorithm~\ref{al0}}.

\subsubsection{Adaptive Bisection Searching Method}
\label{adaptive-based}
The main concept of \ac{abs} is to find a smaller range for feasible $\Gamma_r$, thereby accelerating convergence. 
For convenience, we rewrite the constraint (\ref{gr1}) with an indicator $t$, given by
\begin{equation}
\label{gr2}
    \begin{aligned}
        (1+\Gamma_r^{-1}){\rm tr}\left(\boldsymbol{U}_{l,l}\boldsymbol{R}\right)-\sum_{m=1}^{M}{\rm tr}\left(\boldsymbol{U}_{l,m}\boldsymbol{R}\right) -\sigma_R^2 \geq t,
    \end{aligned}
\end{equation}
where the equivalent noise is represented by $\sigma_R^2=\frac{\sigma_r^2}{\kappa^2 Z^2_0}$. The objective function in (\ref{st3_}) is then converted into finding the maximal $t$ under the fixed $\Gamma_r$, which satisfies the inequality in (\ref{gr2}). 
Subsequently, we employ the indicator $t$ to evaluate the feasibility of the given $\Gamma_r$ for the problem (\ref{st3_}). 

\begin{algorithm}
    \caption{Adaptive Bisection Searching Method}
    \renewcommand{\algorithmicrequire}{\textbf{Input:}}
    \renewcommand{\algorithmicensure}{\textbf{Output:}}
    \label{al1}
    \begin{algorithmic}[1]
        \REQUIRE \leavevmode 
        $\Gamma_c$. \\
        Initial point: $\Gamma_{r,start} = \frac{P_T A_T}{\sigma_R^2}- K\Gamma_c$.
        \WHILE {$1$}
        \STATE Compute the indicator $t$ with fixed $\Gamma_{r,start}$ via (\ref{gr2}).
        \STATE If $t\geq 0$, break; otherwise $\Gamma_{r,start}\leftarrow \Gamma_{r,start}-\Gamma_c$. 
        \ENDWHILE
        \STATE $\Gamma_{r,end} \leftarrow \Gamma_{r,start}+\Gamma_c$.
        \WHILE {$1$}
        \STATE Compute the indicator $t$ with fixed $\Gamma_{r,end}$ via (\ref{gr2}).
        \STATE If $t\leq 0$, break; otherwise $\Gamma_{r,end}\leftarrow \Gamma_{r,end}+\Gamma_c$.
        \ENDWHILE
        \STATE $\Gamma_{r,start} \leftarrow \Gamma_{r,end}-\Gamma_c$.
        \ENSURE \leavevmode 
        $\left[\Gamma_{r,start}, \Gamma_{r,end}\right]$.\\
    \end{algorithmic}
\end{algorithm}

First, for the initial point of the \ac{abs}, we define the maximal potential value of the $\Gamma_r$ as $\Gamma_{r, start}$, satisfying
\begin{equation}
\label{gr_max}
    \begin{aligned}
        \centering
        \Gamma_{r,start} = \frac{P_T A_T}{\sigma_R^2}- K\Gamma_c,
    \end{aligned}
\end{equation}
where $\frac{P_T A_T}{\sigma_R^2}$ is the system's upper bound in the absence of interference-noise, and $K\Gamma_c$ is interference from the multi-user communication. Given $\Gamma_{r,start}$, (\ref{st3_}) can be solved with SDPt3 \cite{toh1999sdpt3}, and the indicator $t$ is computed by (\ref{gr2}). Intuitively, when $t \geq 0$, the actual lower bound of the feasibility solution surpasses the given $\Gamma_{r,start}$, whereas when $t\leq 0$, the actual lower bound of the feasibility solution falls below the given $\Gamma_{r,start}$. Consequently, with the guidance of $t$, a refined range for $\Gamma_r$ can be obtained by iteratively modifying the initial $\Gamma_{r,start}$ with steps $\Gamma_{r,start}\rightarrow\Gamma_{r,start}-\Gamma_c$.

In addition, we define the end point of the potential range as $\Gamma_{r,end} = \Gamma_{r,start}+\Gamma_c$, which should be checked with indicator $t$. Similar to the initial point modification, when $t \geq 0$, the actual upper bound of the feasibility solution surpasses the given $\Gamma_{r,end}$, indicating that $\Gamma_{r,end}\rightarrow\Gamma_{r,end}+\Gamma_c$. After multiple iterations, we obtain the refined dynamic searching range for $\Gamma_r$, denoted as $\left[\Gamma_{r,start}, \Gamma_{r,end}\right]$.
The detailed steps of \ac{abs} are summarized in Algorithm \ref{al1}.

Finally, combined with a confined searching range obtained by the proposed \ac{abs} approach, the \ac{sdr}-based bisection method in Algorithm \ref{al0} is employed to calculate the solution to the problem (\ref{st3_}), with $t$ serving as the solution indicator. Once the search range meets $\Gamma_{r,end}-\Gamma_{r,start}\leq \epsilon_1$ and $t\geq 0$, the iterative algorithm ends to obtain the optimal solution $\boldsymbol{R},\boldsymbol{R}_1,...,\boldsymbol{R}_K$, which we then utilize to calculate the transmit beamformer $\boldsymbol{W}$  via (\ref{W_c}) and (\ref{W_r}).

\subsection{HIS Receive Beamforming with Fixed Transmit Beamforming}
\label{sec: receive design}
For a given transmit beamforming matrix $\boldsymbol{W}$, in order to maximize the minimal sensing \ac{sinr} among all targets, (\ref{P0}) is converted to
\begin{subequations}
\label{P1}
    \begin{align}
    \centering
    \max_{\boldsymbol{Q}} \ \min_{l=1,...,M} &\  \eta_r(\boldsymbol{q}_l;l),\label{P1_obj}\\
    {\rm s.t.} &\ \Vert \boldsymbol{q}_l \Vert_2^2 = 1, l=1,..,M.
    \end{align}
\end{subequations}

The receive beamforming design is still non-convex due to the fractional terms in the objective function. To tackle this problem, we propose a generalized Rayleigh quotient-based method.

First, since the optimization for each receive beamformer $\boldsymbol{q}_l$ w.r.t. the $l$-th target is independent of sensing \ac{sinr} of other targets, (\ref{P1}) can be decoupled into $M$ sub-problems, i.e.,
\begin{subequations}
\label{P2}
    \begin{align}
    \centering
    \max_{\boldsymbol{q}_l} \ &\  \frac{ \boldsymbol{q}_l^H\boldsymbol{A}_l\boldsymbol{q}_l}{\boldsymbol{q}_l^H\left(\boldsymbol{A}-\boldsymbol{A}_l\right)\boldsymbol{q}_l+\frac{\sigma_r^2}{\kappa^2 Z^2_0}}\label{P2_obj},\\
    {\rm s.t.}  &\ \Vert \boldsymbol{q}_l \Vert_2^2 = 1,
    \end{align}
\end{subequations}
where $\boldsymbol{A}_l=\boldsymbol{G}_l\boldsymbol{W}\boldsymbol{W}^{H}\boldsymbol{G}_l^H$ and $\boldsymbol{A}=\sum_{m=1}^{M} \boldsymbol{G}_m\boldsymbol{W}\boldsymbol{W}^{H}\boldsymbol{G}^H_m$.
Then, we have the following lemma based on the generalized Rayleigh quotient \cite{wang2023integrated,zhang2023design}.
\begin{lemma}
    \label{lemma1}
    The objective function in problem (\ref{P2}) can be equivalently written as 
    \begin{equation}
    \label{Ray_pro}
        \begin{aligned}
            \centering
            \max_{\boldsymbol{q}_l} \ &\  \frac{ \boldsymbol{q}_l^H\boldsymbol{B}\boldsymbol{q}_l}{\boldsymbol{q}_l^H\boldsymbol{C}\boldsymbol{q}_l},
        \end{aligned}
    \end{equation}
    where $\boldsymbol{B}=\boldsymbol{A}_l$ and $\boldsymbol{C}=\boldsymbol{A}-\boldsymbol{A}_l+\boldsymbol{I}\frac{\sigma_r^2}{\kappa^2 Z^2_0}$. 
    It is obvious that, for any given beamformer $\boldsymbol{q}_l$, we have the following results: $\boldsymbol{q}_l^H\boldsymbol{B}\boldsymbol{q}_l \geq 0$, $\boldsymbol{q}_l^H\boldsymbol{C}\boldsymbol{q}_l>0$. Therefore, $\boldsymbol{B}\succeq 0$ and $\boldsymbol{C}\succ 0$ are positive semidefinite and positive definite matrices, respectively. Further, both $\boldsymbol{B}$ and $\boldsymbol{C}$ are apparently Hermitian matrices, thereby converting problem (\ref{P2}) into a generalized Rayleigh quotient maximization problem (\ref{Ray_pro}).
    Based on the Rayleigh quotient theory \cite{wang2023integrated,zhang2023design}, the optimal solution of (\ref{Ray_pro}) is given by 
    \begin{equation}
    \label{eigenv}
        \begin{aligned}
            \centering
            \boldsymbol{C}^{-1}\boldsymbol{B}\boldsymbol{v} = \lambda_{max}\left(\boldsymbol{C}^{-1}\boldsymbol{B}\right)\boldsymbol{v},
        \end{aligned}
    \end{equation}
    where $\lambda_{max}\left(\boldsymbol{C}^{-1}\boldsymbol{B}\right)$ is the maximal eigenvalue of $\boldsymbol{C}^{-1}\boldsymbol{B}$, and $\boldsymbol{v}$ is the corresponding eigenvector.
\end{lemma}


Based on Lemma~\ref{lemma1}, the optimal receive beamformer $\boldsymbol{q}_l$ is given by $\boldsymbol{v}$. The corresponding sensing \ac{sinr} is $\lambda_{max}\left(\boldsymbol{C}^{-1}\boldsymbol{B}\right)$.

We summarize the entire process of the proposed \ac{ao}-based joint transmit-receive beamforming for \ac{hisac} system in Algorithm~\ref{al2}. 
\begin{algorithm}
    \caption{AO-Based Joint Transmit-Receive Beamforming for HISAC System}
    \renewcommand{\algorithmicrequire}{\textbf{Input:}}
    \renewcommand{\algorithmicensure}{\textbf{Output:}}
    \label{al2}
    \begin{algorithmic}[1]
        \REQUIRE \leavevmode 
        $\Gamma_c$.
        \REPEAT
        \STATE Determine $(\Gamma_{r,start}, \Gamma_{r,end})$ via Algorithm~\ref{al1}.
        \REPEAT
        \STATE Update $\boldsymbol{R},\boldsymbol{R}_1,...,\boldsymbol{R}_K$ via Algorithm \ref{al0}.
        \UNTIL $\Gamma_{r,end}-\Gamma_{r,start}\leq \epsilon_1$
        \STATE Update $\boldsymbol{W}$ via (\ref{W_c}) and (\ref{W_r}).
        \STATE Update $\boldsymbol{q}_l$ via (\ref{eigenv}).
        \STATE Update the minimal sensing SINR $\Gamma_r^*$.
        \STATE Update $\Gamma_{r,start}\leftarrow\Gamma_r^*$.
        \UNTIL $ \left|\Gamma_r^{*(i)}-\Gamma_r^{*(i-1)}\right| < \epsilon_2$
        \ENSURE \leavevmode 
        $\boldsymbol{W}$, $\boldsymbol{Q}$.
    \end{algorithmic}
\end{algorithm}


Specifically, the transmit beamformers and receive beamformers are updated alternately until a given convergence tolerance between two iterative steps is met, i.e., $ \left|\Gamma_r^{*(i)}-\Gamma_r^{*(i-1)}\right| < \epsilon_2$, where $\Gamma_r^{*(i)}$ and $\Gamma_r^{*(i-1)}$ denote the minimal sensing \ac{sinr} in $(i)$-the and $(i-1)$-th iterations, respectively.



\subsection{Complexity and Convergence Analysis}
\label{sec:conplexity}
\subsubsection{Computational Complexity Analysis} 
The computational complexity of the proposed \ac{ao} algorithm comprises two parts: transmit-side optimization and receive-side update. The transmit-side optimization is formulated as an \ac{sdp}, whose complexity dominates the overall computation. The receive-side update reduces to a generalized Rayleigh quotient problem and admits a closed-form solution. Consequently, each iteration of the proposed algorithm involves solving an \ac{sdp} and then performing a closed-form update. In the following, we provide the computational complexity of each sub-algorithm.
\begin{itemize}
    \item Optimization of transmit beamforming matrices $\boldsymbol{W}_c$, $\boldsymbol{w}_r$ with fixed receive HIS configuration: The \ac{sdp} problem in (\ref{eq_rc}) is solved by the \ac{abs}-\ac{sdr} approach. Specifically, given the potential values of $\Gamma_r$, the number of execution steps of the \ac{abs} $I_o$. Given the solution accuracy $\epsilon_1$, the complexity of using the \ac{sdr} to solve (\ref{eq_rc}) is $\mathcal{O}\left(N^{3.5}\log(1/\epsilon_1)\right)$ \cite{luo2010semidefinite}. Therefore, the computational complexity of transmit beamforming algorithm is $I_o\times \mathcal{O}\left(N^{3.5}\log(1/\epsilon_1)\right)$.

    \item Optimization of receive HIS configuration with fixed transmit beamforming matrices: Since we suggest the generalized Rayleigh quotient algorithm to address (\ref{P2}), the complexity of this algorithm is $\mathcal{O}\left( N^3\right )$.
\end{itemize}
    
As a result, the computational complexity of the proposed alternating algorithm for each iteration is $I_o\times \mathcal{O}\left(N^{3.5}\log(1/\epsilon_1)\right)+\mathcal{O}\left( N^3\right )$.   

\subsubsection{Convergence Analysis}
Due to the non-convex nature of the joint transmit-receive beamforming problem, global optimality of Algorithm~\ref{al2} cannot be formally guaranteed. However, the proposed \ac{ao} algorithm ensures convergence to a stationary point, as each subproblem yields either a convex formulation or a closed-form solution, and guarantees that the objective value does not decrease in each iteration.
On the receiver side, the HIS configuration is optimized using a generalized Rayleigh quotient, which yields a closed-form solution without iterative updates. Therefore, convergence analysis primarily focuses on the transmit-side HIS optimization and the overall \ac{ao} process.
For the transmit-side configuration, we adopt an adaptive algorithm based on the \ac{sdr} framework, termed \ac{abs}-\ac{sdr}. The algorithm iteratively adjusts the search interval $\left[\Gamma_{r,start}, \Gamma_{r,end}\right]$ according to the achieved sensing \ac{sinr}. If the \ac{sinr} improves, the new configuration is accepted and the search region is tightened. Otherwise, the algorithm reverts to the previous configuration. 
Since both sensing and communication \ac{sinr} are inherently bounded due to physical limitations such as total transmit power, this ascent property guarantees the convergence of the \ac{abs}-\ac{sdr} algorithm.
As both subproblems exhibit the ascent property and are constrained within a limited power constraint, the overall \ac{ao} algorithm is guaranteed to converge.


	\section{Numerical Results}
	\label{sec:Sims}
	
In this section, we provide numerical results to evaluate the performance of our proposed \ac{hisac} system and validate the effectiveness of the proposed algorithms. 

\subsection{Simulation Setup}
\label{subsec:setup}

\begin{figure}[htbp]
    \centering
    \includegraphics[width=0.9\linewidth]{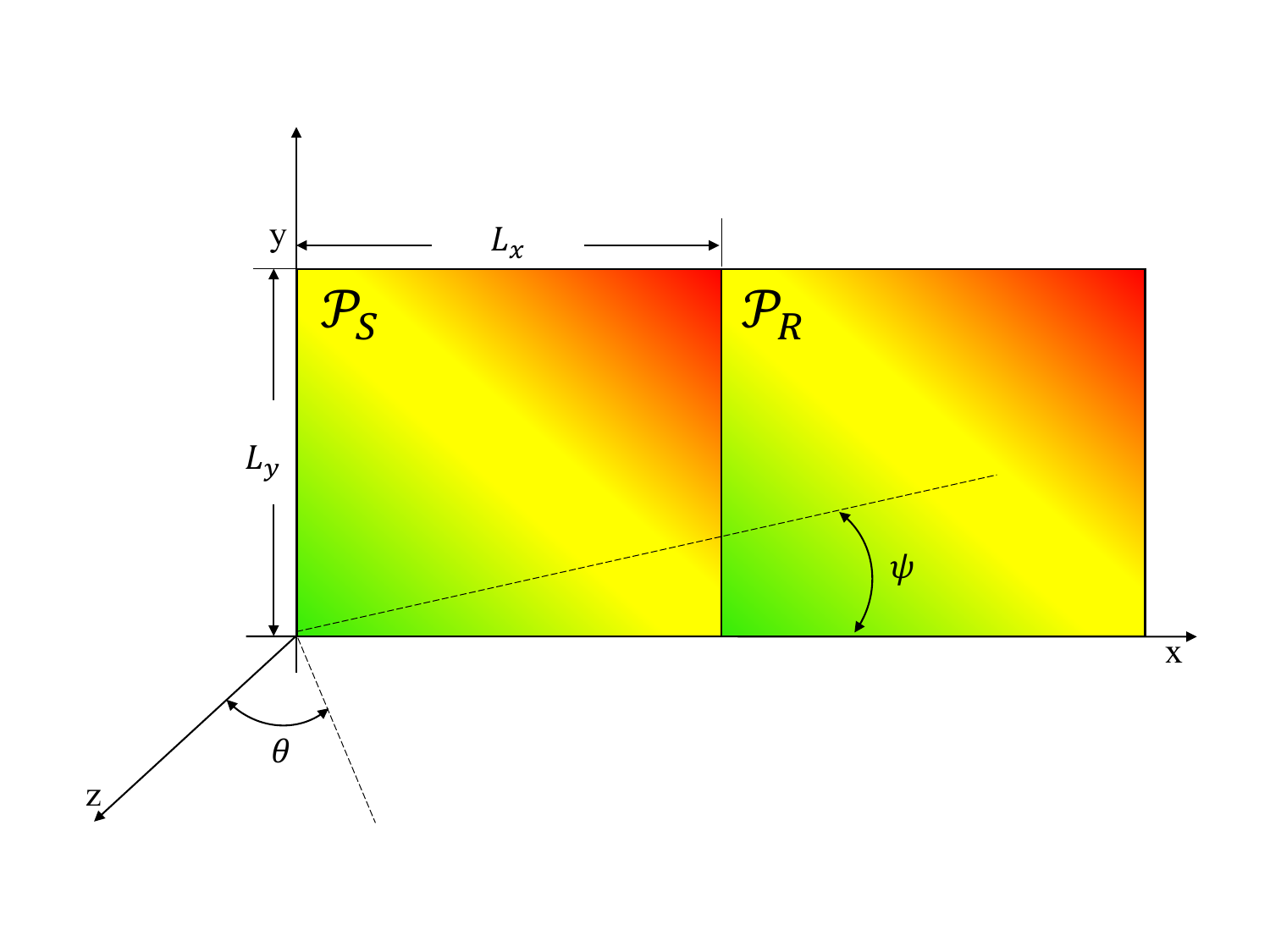}
    \caption{Illustration of the proposed HISAC simulation scenario.}
    \label{fig:geo_sim}
\end{figure}

Without specified otherwise, the simulation parameters are summarized in Table~\ref{tab:hisplat}, and the geometry of the simulation scenario is illustrated in Fig.~\ref{fig:geo_sim}. 
The $\mathcal{P}_S$ and $\mathcal{P}_R$ are located in the XOY plane adjacently. Simultaneously, all communication users and radar targets are located on the far-field radiation region of the \ac{his}, represented by polar coordinates.
Additionally, to underscore the superiority of the proposed \ac{hisac} system, we utilize the discrete array counterpart, where antennas are deployed with half-wavelength spacing, as a baseline. 
Note that the transmit power $P_T$ is computed based on the continuous-aperture current model, where the unit is expressed in mA$^2$ rather than in watt \cite{zhang2023pattern}.
Moreover, the truncation order is not optimized on a per-scenario basis. Instead, it is determined by the physical aperture size and the carrier wavelength, which jointly limit the wavenumber bandwidth of the radiated field. Once the aperture is fixed, the truncation order can be selected according to (\ref{first_zero}), and remains unchanged across various communication and sensing scenarios.

\begin{table}[htbp]
    \centering
    \caption{The basic simulation parameters}
    \begin{tabular}{c|c}
    \hline \hline
        Parameters & Value \\
        \hline
        Center frequency & $2.4$~GHz \\
        Aperture size &  $ A_T = 0.5\times 0.5$~m$^2$\\
        Transmit power budget & $P_T = 100$~mA$^2$\\
        Position of the $1$-th user & $(30^\circ, 180^\circ)$ \\
        Position of the $2$-th user & $(30^\circ,270^\circ)$ \\
        Position of the target $1$ & $(30^\circ, 90^\circ)$\\
        Position of the target $2$ & $(30^\circ, 45^\circ)$\\
        $\Gamma_c$ & 5 dB \\
        \hline \hline
    \end{tabular}
    \label{tab:hisplat}
\end{table}

\begin{remark}
    \label{dis_rad}
    Compared to the \ac{his}, we assume the discrete-array-based system consists of discrete patches (isotropic source) with half-wavelength inter-distance. The aperture size of the discrete array remains the same as the \ac{his}, allowing $D=DxDy=\frac{L_x}{\lambda/2}\frac{L_y}{\lambda/2}$ antennas to be optimized.
According to \cite{balanis2016antenna}, the efficient aperture size of the isotropic source is given by $e_a=\frac{\lambda^2}{4\pi}$. Therefore, the maximal radiation power of the discrete array with $D$ elements can be calculated as $\frac{P_T A_T}{\pi}.$
Based on this setup, compared to the \ac{hisac} system, the round-trip performance degradation of the discrete array counterpart is $\left(\frac{e_a D}{L_xL_y}\right)^2=\frac{1}{\pi^2}$.
\end{remark}


\subsection{Performance Evaluation}
\label{subsec:converg}
\subsubsection{Algorithm Convergence}

\begin{figure}
	\centering  
	\vspace{-0.35cm} 
	\subfigtopskip=2pt 
	\subfigbottomskip=2pt 
	\subfigcapskip=-5pt 
	\subfigure[]{
		\label{fig:convergence_aia}
		\includegraphics[width=0.95\linewidth]{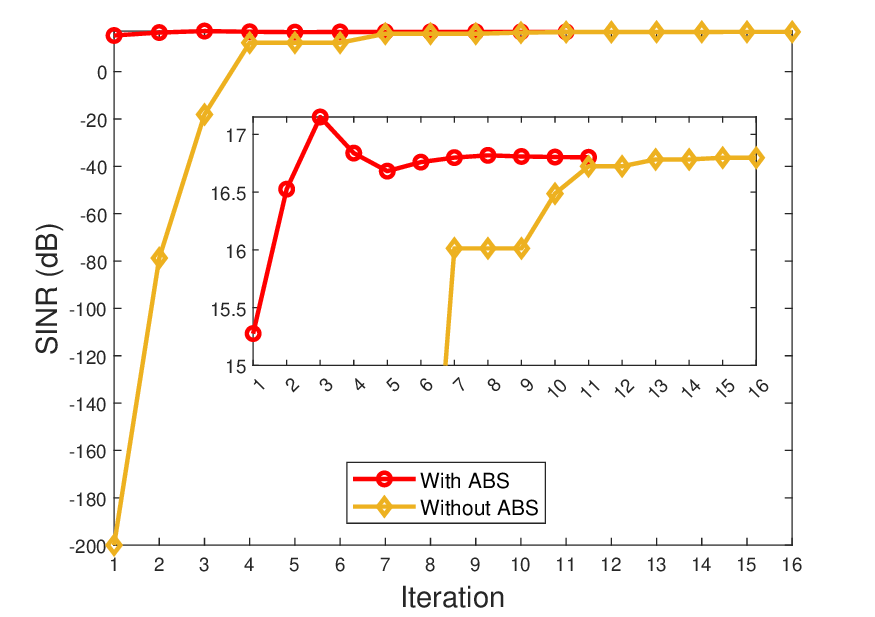}}
  
	\subfigure[]{
		\label{fig:convergence_bs}
		\includegraphics[width=0.95\linewidth]{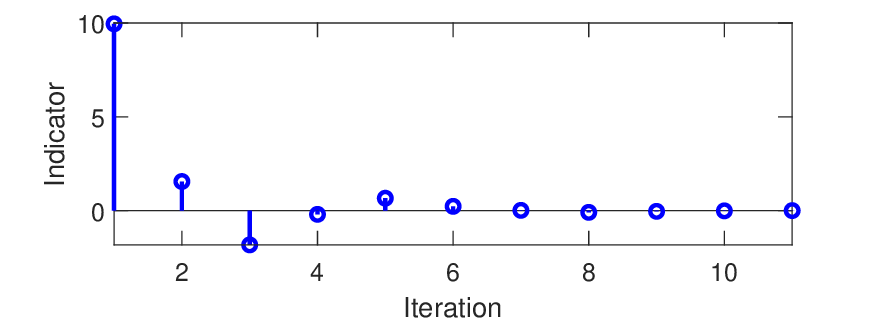}}
	\caption{Illustration of algorithm convergence. (a) Convergence of Algorithm~\ref{al0} with \ac{abs} and without \ac{abs}. (b) Convergence of the indicator $t$.}
	\label{fig:convergence}
\end{figure}

First, we study the convergence of the \ac{ao}-based approach in Algorithm~\ref{al2} and the \ac{sdr}-based method in Algorithm~\ref{al0} with \ac{abs} acceleration in Algorithm~\ref{al1}. In particular, Algorithm~\ref{al2} converges within two steps, demonstrating its efficacy and robustness. Fig. \ref{fig:convergence_aia} presents the convergence trends of the \ac{sdr}-based method with \ac{abs} and without \ac{abs} acceleration. 
We can see that, with the proposed \ac{abs} in Algorithm~\ref{al1}, the search range of $\Gamma_r$ is shrunk, accelerating the convergence of the \ac{sdr}-based bisection search in Algorithm~\ref{al0}.
Specifically, under a given convergence tolerance, the proposed algorithm converges within $11$ steps, while the algorithm without \ac{abs} requires another $6$ steps to reach its convergence.
To evaluate the impact of the indicator $t$ on the proposed algorithm's convergence, Fig.~\ref{fig:convergence_bs} shows the convergence trend of the indicator $t$.
As can be seen, the indicator $t$ converges to $0$, implying the optimal solution to (\ref{st3_}) is found. Specifically, when $t>0$, the potential value $\Gamma_r$ is feasible but not optimal; when $t<0$, the potential value $\Gamma_r$ is infeasible. If and only if $t\to 0$, the solution of (\ref{st3_}) converges to the optimal value. Therefore, the minimal sensing \ac{sinr} oscillates and converges to the maximal value after multiple iterations.


\subsubsection{Multi-User Single-Target Scenario}
\label{multi-user single target}
We evaluate the performance gain of our proposed \ac{hisac} system with its discrete array counterpart and verify the effectiveness of the \ac{ao} algorithm by visualizing the transmit beams. 
For comparison, the upper bound of sensing performance in the \ac{hisac} system is defined as the sensing \ac{sinr} without communication interference, which can be calculated by $\frac{P_TA_T}{\sigma_R^2}$ via (\ref{gr_max}).

\begin{figure}
    \centering
    \includegraphics[width=0.95\linewidth]{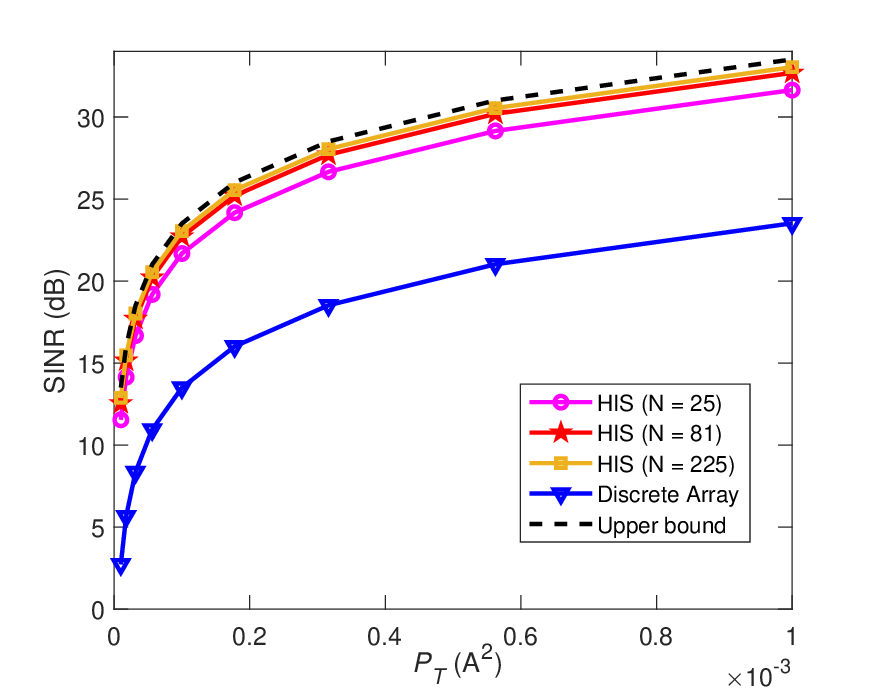}
    \caption{Sensing SINR versus the power budget $P_T$.}
    \label{fig:sinrvspt}
\end{figure}

In Fig. ~\ref{fig:sinrvspt}, we evaluate the sensing \ac{sinr} versus the maximal transmit power $P_T$  for different \ac{isac} configurations. 
We observe that our proposed \ac{hisac} system outperforms the discrete array counterpart in the considered range of power budget. 
Moreover, with sufficiently large transmit power, the proposed system exhibits a $9.5$~dB sensing \ac{sinr} improvement compared to the discrete-array-based \ac{isac} system. The reason is that, under the same physical aperture size, the efficient size of the discrete array is smaller than that of \ac{his}, leading to insufficient power radiation and receiving. 
As stated in Remark~\ref{dis_rad}, the round-trip radiation power degradation compared to \ac{his} is $\pi^2$, resulting in the sensing \ac{sinr} undergoing $9.94$~dB. Therefore, with the fulfilling capture of the \ac{his} radiation, the performance gain of the \ac{hisac} system compared to the discrete array will stabilize around $\pi^2\rightarrow9.94$~dB.

\begin{figure}[htbp]
    \centering
    \includegraphics[width=0.95\linewidth]{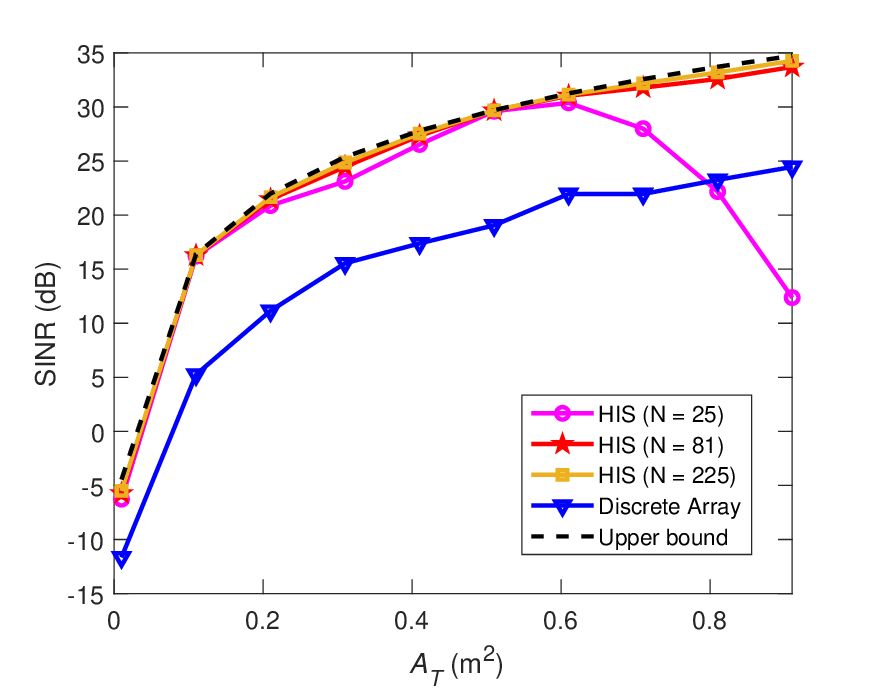}
    \caption{Sensing SINR versus the aperture size $A_T$ of the transmit \ac{his}.}
    \label{fig:sinrvsat}
\end{figure}

To explore the impact of aperture size on the \ac{hisac} system, we evaluate the sensing \ac{sinr} versus the aperture area $A_T$ in Fig. \ref{fig:sinrvsat}, where the transmit \ac{his} always remains a square with $L_x = L_y =\sqrt{A_T}$.
As can be seen, the radar performance improves with the increase of the aperture size, and the sensing \ac{sinr} of our proposed \ac{hisac} system exhibits a progressive enhancement as the Fourier expansion order $N$ increases, outperforming its discrete array counterpart even when $N$ is moderately large.
However, if the Fourier expansion order is too small to adequately capture the \ac{his} radiation characteristics, the radar performance degrades significantly in the large-aperture region and may fall below that of the discrete array counterpart.

\begin{figure}
	\centering  
	\vspace{-0.35cm} 
	\subfigtopskip=2pt 
	\subfigbottomskip=2pt 
	\subfigcapskip=-5pt 
	\subfigure[]{
		\label{fig:trans_beam1}
		\includegraphics[width=0.85\linewidth]{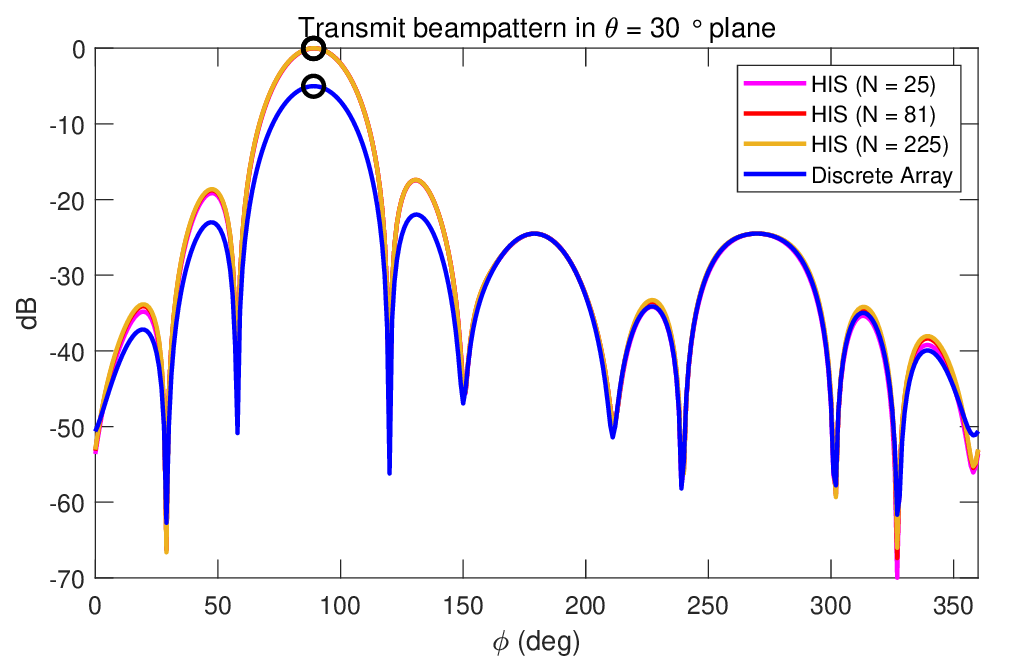}}
  
	\subfigure[]{
		\label{fig:rece_beam1}
		\includegraphics[width=0.85\linewidth]{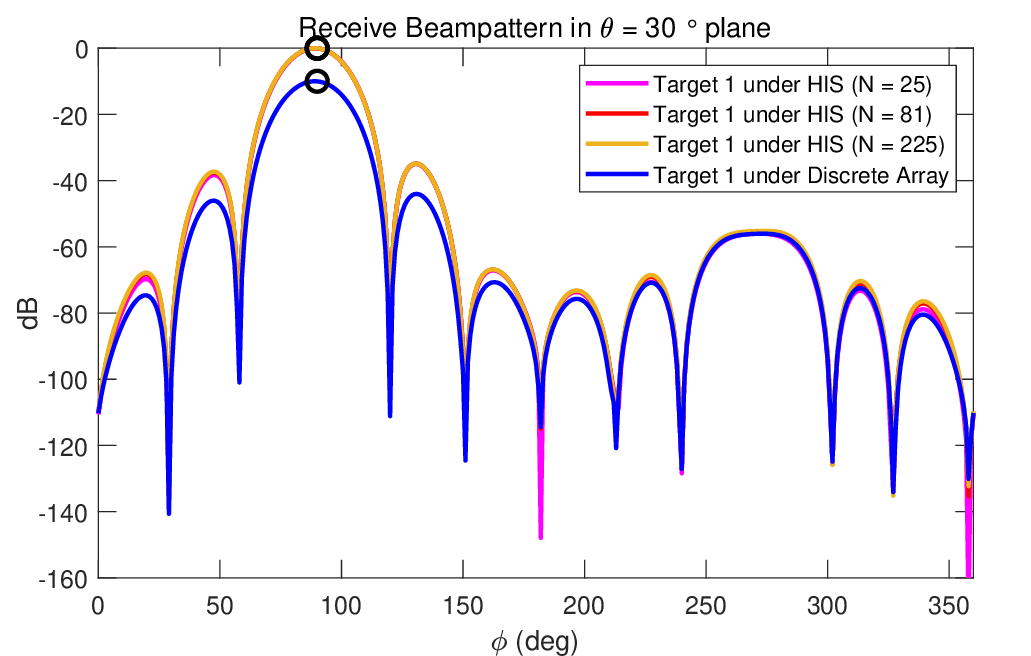}}
	\caption{HIS beampatterns for target $1$. (a) Transmit beampattern. (b) Receive beampattern.}
	\label{fig:beam_mt1}
\end{figure}

In Fig.~\ref{fig:trans_beam1} and Fig.~\ref{fig:rece_beam1}, we illustrate the transmit and receive beampatterns of the \ac{his} and that of the discrete array at $\theta = 30^\circ$ plane, respectively. Based on these normalized beampatterns, we study the beampattern gain between the \ac{his}-based and discrete-array-based system.
As can be seen, in the transmit phase, as the Fourier expansion order $N$ increases, the beamwidth of the \ac{his}'s transmit beampattern tightens, where the peak power reaches its maximum until the $N$ is sufficient for characterizing the \ac{his} radiation. Meanwhile, the peak value of the transmit beampattern based on the discrete array demonstrates $4.9$~dB degradation compared to \ac{his}, which agrees with $\pi$. In addition, for the receive beampattern w.r.t. target $1$, the peak value of the \ac{his}'s beampattern outperforms that of the discrete array with a $9.7$~dB enhancement. 
As a result, the sensing \ac{sinr} for target $1$ can be significantly improved after the receive beamforming under the proposed \ac{hisac} transceiver architecture. 
The reason is that both the \ac{his}'s transmit and receive beamformers are optimized at the receiver side, leading to the overall sensing \ac{sinr} improvement in our proposed architecture.

\begin{figure}
    \centering
    \includegraphics[width=0.95\linewidth]{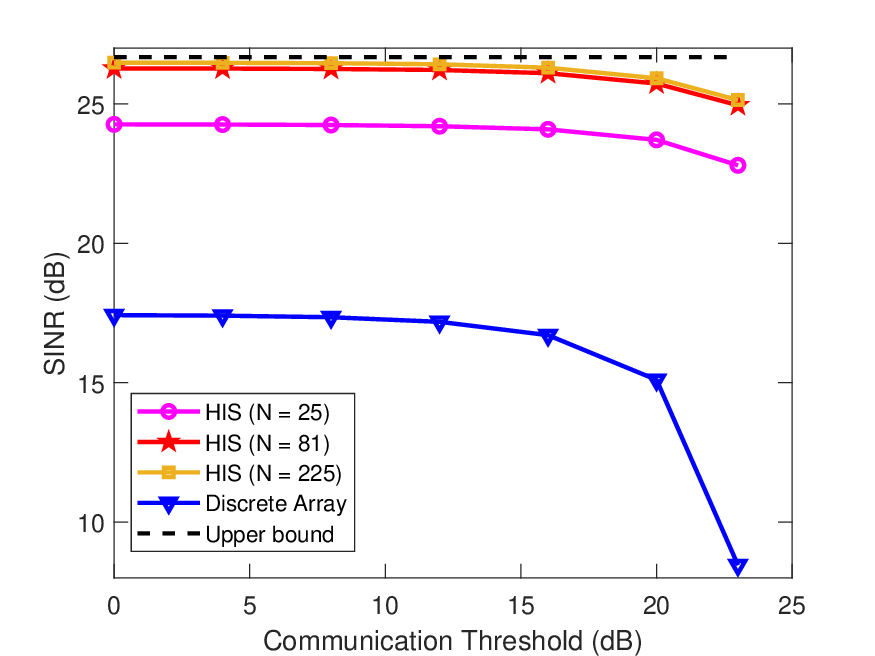}
    \caption{Sensing SINR versus communication SINR threshold $\Gamma_c$.}
    \label{fig:maxsinr_com}
\end{figure}

In Fig.~\ref{fig:maxsinr_com}, we explore the sensing \ac{sinr} under varying communication \ac{sinr} requirements. Specifically, we evaluate sensing \ac{sinr} versus different communication threshold $\Gamma_c$. The physical aperture size is $A_T = 0.6\times 0.6$~m$^2$ to demonstrate the superior sensing \ac{sinr} difference between Fourier expansion order $N$ and its discrete array counterpart.
A clear trade-off between sensing and communication is presented, in which the sensing \ac{sinr} decreases as the communication \ac{sinr} requirement increases. In addition, we observe that our proposed \ac{hisac} system shows robustness compared to the discrete array system, possessing more significant sensing performance gain with higher communication \ac{sinr} threshold. Specifically, as $\Gamma_c$ increases, the sensing \ac{sinr} gain of the \ac{hisac} system over the discrete array-based \ac{isac} system increases from $7.4$~dB to $15.6$~dB, demonstrating the advantages of incorporating \ac{his} into \ac{isac} systems.

\subsubsection{Multi-User Multi-Target Scenario}
\label{multi-user multi-target}

\begin{figure}
    \centering
    \includegraphics[width=0.95\linewidth]{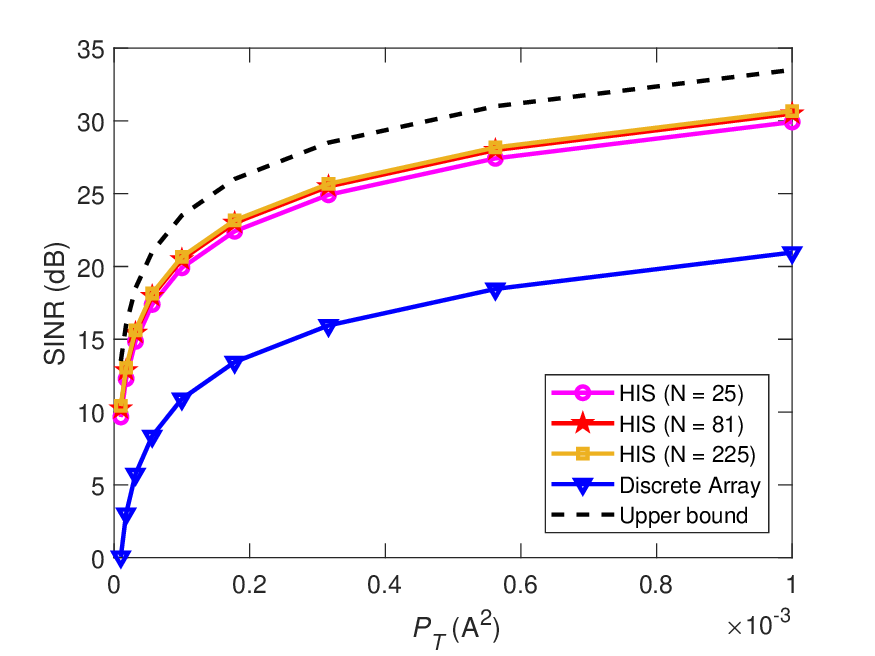}
    \caption{Minimal sensing SINR versus the power budget $P_T$.}
    \label{fig:sinrvspt_mt}
\end{figure}

To evaluate the scalability and effectiveness of the proposed HISAC framework, we provide comprehensive simulations under a multi-user and multi-target scenario. 
An additional radar target positioned at $(30^{\circ}, 45^{\circ})$ is introduced in the multi-user multi-target scenario. In Fig.~\ref{fig:sinrvspt_mt}, we compare the minimal sensing \ac{sinr} among radar targets for different Fourier expansion order $N$ versus $P_T$. 
As observed, due to inter-target interference, the sensing \ac{sinr} undergoes a $3$ dB degradation, compared to the interference-free upper bound. Nevertheless, our proposed HISAC system still outperforms its discrete array counterpart with $9.7$~dB sensing \ac{sinr} enhancement. 
The insight here is that the number of radar targets will not affect the sensing \ac{sinr} enhancement between the \ac{his} and the discrete array. This is because the radiation power of both architectures is independent of the number/directions of targets. Therefore, when the Fourier expansion order $N$ is sufficient, our system can still achieve power gain around $\pi^2$.

\begin{figure*}
	\centering  
	\vspace{-0.35cm} 
	\subfigtopskip=2pt 
	\subfigbottomskip=2pt 
	\subfigcapskip=-5pt 
	\subfigure[]{
		\label{fig:trans_beam}
		\includegraphics[width=0.9\linewidth]{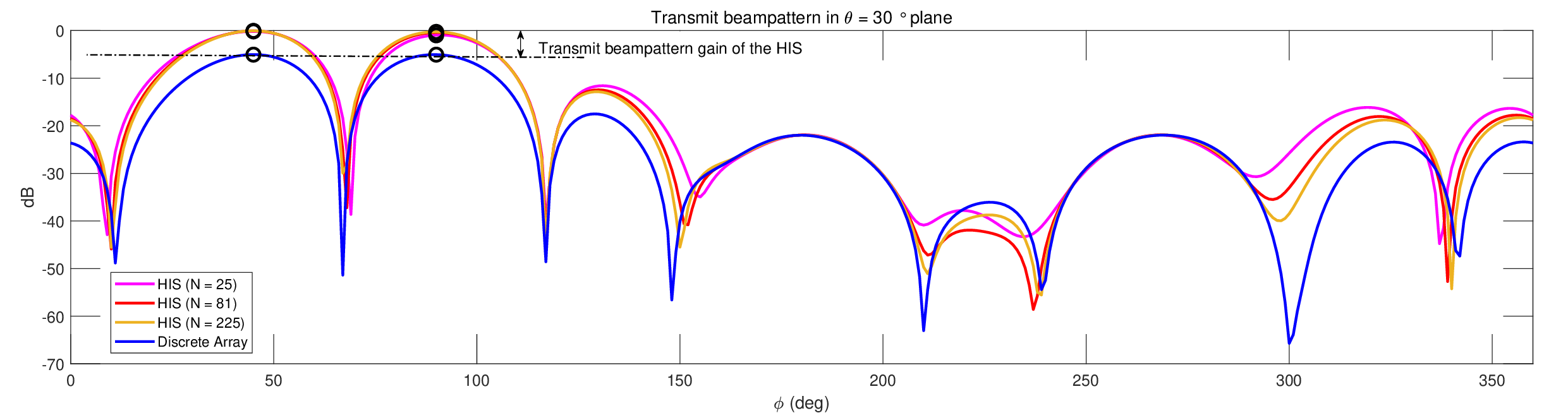}}
  
	\subfigure[]{
		\label{fig:rece_beam}
		\includegraphics[width=0.9\linewidth]{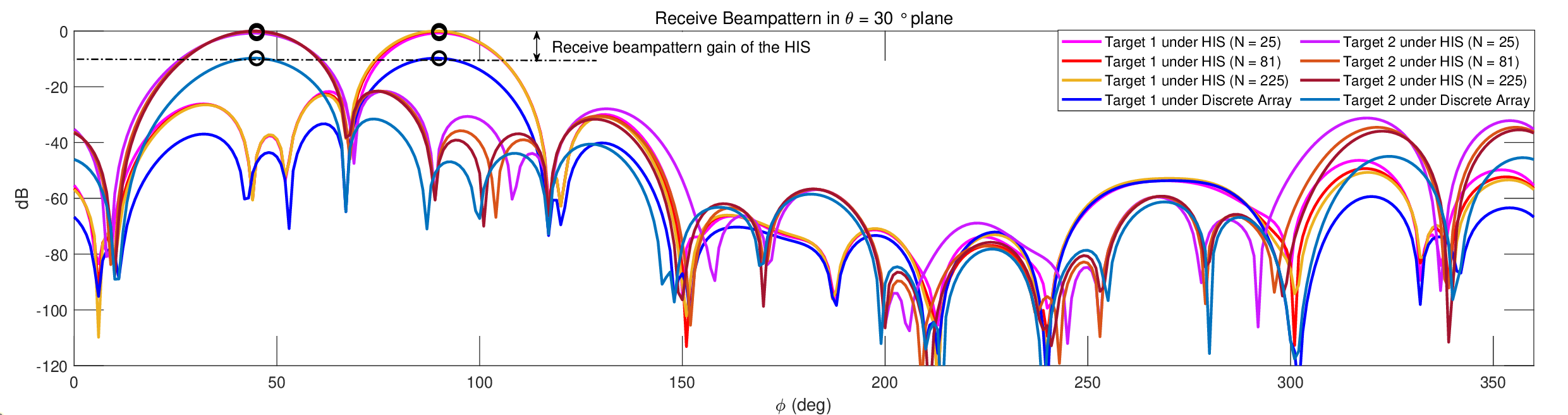}}
	\caption{HIS beampatterns for multi-target. (a) Transmit beampattern. (b) Receive beampattern.}
	\label{fig:beam_mt}
\end{figure*}

In Fig.~\ref{fig:beam_mt}, we illustrate the normalized transmit and receive beampatterns at $\theta=30^\circ$ plane to demonstrate the impact of multi-target sensing on the minimal sensing \ac{sinr}. 
As observed in Fig.~\ref{fig:trans_beam}, the main lobe of the \ac{his} transmit beampattern is split into two main lobes with the same peak level. Compared to the discrete array, the transmit beampattern of \ac{his} exhibits a $4.9$~dB enhancement in each direction of targets. As can be seen in Fig.~\ref{fig:rece_beam}, the receiving beamforming strengthens the beampattern's peak values in the directions of targets while suppressing those of the clutters (targets located in other directions) deeply. Specifically, for the receive beamforming w.r.t. target $1$, the direction w.r.t. target $1$ is enhanced while the direction w.r.t. target $2$ is suppressed and vice versa. Note that the peak value of \ac{his} still maintains an enhancement around $9.7$~dB compared to the discrete array, regardless of the directions of the receive beamforming. 
Therefore, our proposed algorithm is efficient in addressing the multi-user multi-target scenario and indicates that the sensing \ac{sinr} improvement is attributed to the efficient aperture and is independent of the directions of targets.

\begin{figure}[htbp]
    \centering
    \includegraphics[width=0.95\linewidth]{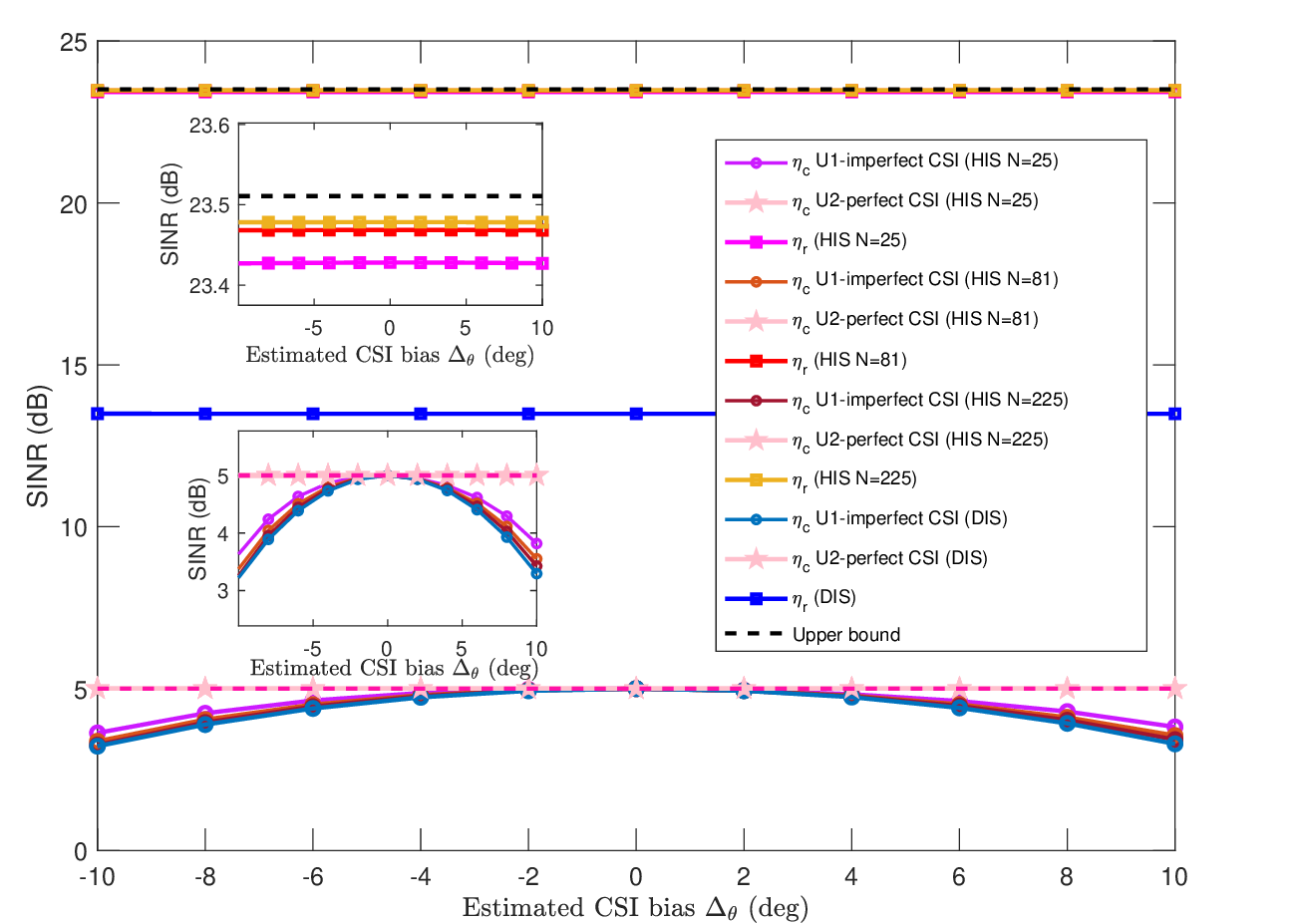}
    \caption{SINRs for sensing and communication versus imperfect CSI bias.}
    \label{fig:imperCSI1}
\end{figure}
To further evaluate the robustness of the proposed HISAC system under channel uncertainty, we conduct simulations under imperfect CSI conditions. The angular deviation $\Delta_\theta$ represents the mismatch between the estimated and actual angle of arrival/departure for \textit{User 1}, thereby capturing CSI inaccuracy.
We compare the sensing and communication SINR across all radar targets and communication users for both HIS-based and discrete-array-based systems. 
As shown in Fig.~\ref{fig:imperCSI1}, the results reveal a distinct trade-off between the Fourier truncation order $N$ of the HIS and its robustness to CSI error. 
Specifically, higher-order HIS configurations achieve better radiation fidelity and stronger beamforming gain under ideal CSI, but become more sensitive to imperfect CSI.
Nevertheless, all HIS configurations consistently outperform their discrete-array-based counterparts, even under moderate CSI error. Results demonstrate that the efficient aperture utilization of HIS enables stable SINR performance for both sensing and communication tasks.

	\section{Conclusion and Discussion}
	\label{sec:Conclusions}

This paper introduced a novel concept to ISAC systems by proposing the HISAC system. 
Through the continuous-aperture array fabrication of both transmitter and receiver, the HISAC system presented unique challenges in design.
To address these, we developed a Fourier-based continuous–discrete transformation that maps the continuous pattern into a finite-dimensional wavenumber domain, enabling compatibility with traditional discrete-array ISAC frameworks
Based on this reformulation, a joint transmit-receive beamforming problem was formulated to balance multi-target sensing performance with multi-user communication requirements. An AO-based algorithm was developed to efficiently tackle the non-convex optimization with coupled variables. 
Numerical results demonstrated superior performance compared to traditional discrete-array-based ISAC systems, achieving significantly enhanced sensing capabilities while maintaining predetermined communication performance levels.

\subsection{Discussion}
\label{discussion}
\subsubsection{Hardware Implementation Challenges in HIS Systems}

This work focuses on the algorithmic and theoretical modeling of continuous-aperture \ac{his}. However, practical implementation introduces several hardware-level challenges that must be addressed before real-world deployment. 
The proposed \ac{hisac} framework assumes a fully-digital \ac{his} architecture, in which each beam, corresponding to a Fourier-domain coefficient, is driven by a dedicated \ac{rf} chain. However, implementing such a system in practice entails significant hardware complexity, particularly due to the delicate \ac{rf} chains.
To bridge the gap between theoretical models and practical
systems, future implementations may consider hybrid \ac{his} architectures, which offer a trade-off between flexibility and complexity. In addition, hardware imperfections such as phase quantization, nonlinearities, and finite control resolution should be carefully modeled to understand their impact on system performance. These non-idealities primarily distort the effective \ac{his} beampattern and can be incorporated into the proposed framework through uncertainty-aware constraints or robust optimization formulations.

\subsubsection{Scalability in Multi-HIS and Large-Scale Networks}

The proposed \ac{hisac} system already adopts a dual-\ac{his} configuration, in which separate transmit and receive \acp{his} are deployed to support full-duplex \ac{isac} functionality. However, extending the framework to cell-free \ac{mimo} or multistatic sensing, which requires the cooperation among multiple base stations/\acp{his}, has yet to be explored.
In such distributed settings, inter-\ac{his} synchronization, mutual interference suppression, and coordinated beam scheduling become essential. As the number of cooperative \ac{his} nodes increases, the dimensionality of control variables grows accordingly, which substantially increases the computational complexity of joint optimization.
For large-scale deployments involving multiple \acp{his}, future research may investigate low-rank structural modeling, learning-assisted beam coordination strategies, and distributed optimization algorithms. 
These approaches can facilitate scalable and adaptive system operation while maintaining global performance constraints.

\subsubsection{System-Level Design Based on the HIS representation}

Under dual-\ac{his} full-duplex operation, \ac{si} arises from spectral coupling between transmit and receive excitations, which is typically overlooked in conventional \ac{isac} studies. 
Although the proposed Fourier-domain representation provides a promising framework for characterizing such coupling, its integration with dedicated \ac{si} suppression mechanisms remains an open problem.
Moreover, the current \ac{isac} signal model is formulated based on stochastic signal design via covariance optimization. Extending this framework toward deterministic sensing waveform design is another important direction for future investigation.
At the system level, additional performance metrics, such as communication capacity, energy efficiency, and real-time computational latency, will also be explored to achieve a more comprehensive evaluation of practical performance.

\vspace{-0.2cm}
\begin{appendix}
	\numberwithin{proposition}{subsection} 
	\numberwithin{lemma}{subsection} 
	\numberwithin{corollary}{subsection} 
	\numberwithin{remark}{subsection} 
	\numberwithin{equation}{subsection}

\subsection{Proof of Remark~\ref{remark} }
\label{remark1}

We now justify the i.i.d. Gaussian model adopted in the discrete domain. Specifically, we assume the receive-side noise is a spatially white circularly symmetric complex Gaussian random field satisfying \cite{Sanguinetti2022wavenumber}
\begin{equation}
\label{eqv1}
    \begin{aligned}
        \centering
        \mathbf{E}\left(v_r(p')v_r^*(p'') \right)=\sigma_r^2 \delta\left(p'-p''\right),~p',p''\in D,
    \end{aligned}
\end{equation}
where $\delta(\cdot)$ is the Dirac delta function. Let $\left\{\boldsymbol{\Phi}_n(p')\right\}_{n=1}^N$ be the truncated orthonormal Fourier basis. 
The $n$-th Fourier coefficient of the noise is defined as
\begin{equation}
\label{eqv2}
    \begin{aligned}
        \centering
        v_{r,n} = \int_D v_r(p')\boldsymbol{\Phi}_n(p') \mathrm{d}p',~n=1,...,N.
    \end{aligned}
\end{equation}
Collecting $v_{r,n}$ into vector $\boldsymbol{v}_r$, with the the orthogonality of $\boldsymbol{\Phi}_n(\cdot)$, we obtain
\begin{equation}
\label{eqv3}
    \begin{aligned}
        \centering
        &\mathbf{E}\left(\boldsymbol{v}_r\boldsymbol{v}_r^H \right)=\\
        &\left[\int_D\int_D \mathbf{E}\left(v_r(p')v_r^*(p'') \boldsymbol{\Phi}_n(p')\boldsymbol{\Phi}_n^*(p'')\right)\mathrm{d}p'\mathrm{d}p''\right]_{n,n'}\\
        &=\sigma_r^2 \boldsymbol{I}_N.
    \end{aligned}
\end{equation}
Therefore, $\boldsymbol{v}_r \sim\mathcal{CN}\left(0,\sigma_r^2\boldsymbol{I}_N\right)$, which validates the noise model in (23).

This concludes the proof.

\end{appendix}

	\bibliographystyle{IEEEtran}
	\bibliography{IEEEabrv,main}

\end{document}